\documentclass[%
 reprint,
 aps,
 pre,
 showkeys,
 floatfix
]{revtex4-2}

\usepackage{graphicx}
\usepackage{dcolumn}
\usepackage{bm}
\usepackage{color}
\usepackage{xcolor}
\usepackage[normalem]{ulem}
\usepackage{amssymb}
\usepackage{amsmath}
\usepackage{hyperref}
\usepackage{booktabs}

\begin{document}

\preprint{APS/123-QED}

\title{Inferring Latent Geometries in Weighted Spatio-Functional Networks}

\author{Gabriele Cerioli\textsuperscript{1}, Adamo Cerioli\textsuperscript{2, *}}

\affiliation{
 \textsuperscript{1}Département de Physique, École Normale Supérieure, Université PSL, 45 rue d'Ulm, 75230 Paris Cedex 05, France
}
\affiliation{
 \textsuperscript{2}Department of Mathematics, Physics and Computer Science, University of Parma, Parco Area delle Scienze, 7/A, 43124, Parma, Italy
}
\affiliation{
 \textsuperscript{*}Corresponding author: adamo.cerioli@unipr.it
}

\begin{abstract}
\noindent
Spatial constraints and functional demands represent two deeply intertwined drivers across complex networked systems: geographic embedding imposes metric wiring costs, while functional requirements drive topological routing and interaction patterns. Despite their mutual feedback, empirical systems rarely conform to idealized dichotomies of pure planar grids or unconstrained relational topologies, while existing latent-space models remain computationally bottlenecked by Markov chain Monte Carlo sampling on small, unweighted graphs. Here, we formulate a scalable, generative statistical framework to solve the inverse problem of spatio-functional inference in weighted networks. By coupling explicit metric spaces with an unobserved latent functional geometry, we jointly reconstruct hidden node coordinates and infer an explicit coupling parameter $\lambda \in [0, 1]$, which quantifies the relative connectivity propensity explained by latent functional affinity conditioned on metric distance decay. Our deterministic, unsupervised pipeline combines a physics-informed spectral initialization on spatial residuals with hybrid first- and quasi-Newton optimization. Applying this framework across biological, technological, and mobility networks reveals a continuous spectrum of spatio-functional topologies. In the \textit{Drosophila melanogaster} whole-brain connectome, the model resolves an architectural continuum transitioning from metric-constrained sensory circuits to functionally specialized associative and chemo-affinity neuropils. In the continental European power grid, stress tests show that line pruning preserves a strictly metric regime ($\hat{\lambda} \approx 0$), whereas topological rewiring drives a sharp increase in $\hat{\lambda}$, revealing that high component connectivity can mask structural spatial breakdown. Finally, tracking the global air transportation network during the COVID-19 pandemic captures dynamic regime shifts toward localized, metric connectivity. More broadly, these results illustrate how disentangling metric distance decay from latent functional affinity offers a unified perspective to investigate networks across diverse empirical domains.
\end{abstract}


\maketitle

\noindent

Throughout history, collective human dynamics have continuously shaped the topology of physical environments, giving rise to complex network architectures tailored to functional demands, spanning from local pedestrian pathways and university campuses to regional transit grids and international aviation corridors \cite{BARTHELEMY20111, batty2013newscience}. Conversely, the physical configuration of spatial networks shapes social interactions and broader demand patterns by facilitating or constraining encounters, communication, and resource exchange across geographic space \cite{hillier1989social, onnela2011geographic}. Rather than operating as decoupled regimes, spatial constraints and functional demands exist within a perpetual feedback loop: spatial networks cannot be fully understood without their underlying functional drivers, just as functional connectivity remains fundamentally anchored to physical geometry and infrastructural affordances \cite{gonzalez2008understanding, liben2005geographic}.

To characterize these interconnected systems, network science provides an established mathematical foundation \cite{newman2003structure, boccaletti2006complex}. Within this framework, purely functional systems, canonically exemplified by social networks, are modeled as relational ties exhibiting small-world topologies unconstrained by physical space \cite{wasserman1994social, watts1998collective}. In contrast, purely spatial networks embed nodes in coordinate space, where geometric wiring costs enforce bounded degrees and sharp distance decay \cite{crucitti2006centrality, blanchard2009mathematical}.

In empirical networks, however, this dichotomy collapses. Geographic proximity anchors functional interactions \cite{herrera2015anatomy, cagney2020urban}, while physical infrastructures introduce non-local shortcuts to satisfy long-range demands \cite{guimera2005worldwide, xu2019quantifying}. Consequently, real-world networks inhabit a continuous spatio-functional spectrum that standard topological heuristics fail to capture.

While structural gravity models absorb non-spatial affinities into node-specific fixed effects \cite{anderson2003gravity, behrens2012dual}, existing latent space formulations that incorporate geographic covariates \cite{ward2013gravity} or evaluate sociodemographic distance against physical geography \cite{nandy2025socio} remain fundamentally bottlenecked by Markov chain Monte Carlo (MCMC) sampling \cite{salter2013variational}, restricting their deployment to small graphs or areal units. Here, we model observed weighted connectivity as a dual-channel generative superposition of geographic distance decay and latent topological proximity \cite{butts2011spatial, xu2022beyond}, governed by an explicit spatio-functional coupling parameter $\lambda \in [0, 1]$ that isolates the relative contribution of latent functional affinity beyond geographic distance decay.

To resolve the non-convex landscape and rotational gauge symmetries of the latent geometry, we formulate a deterministic inference pipeline that bypasses the convergence instabilities of stochastic sampling. Latent coordinates are first initialized via spectral decomposition of the connection residuals unexplained by geography \cite{golub1987generalization, gallagher2021spectral}. These positions are subsequently refined through a two-stage scheme pairing an Adam warm-up \cite{kingma2014adam} with limited-memory Broyden--Fletcher--Goldfarb--Shanno (L-BFGS) quasi-Newton optimization \cite{liu1989limited, sivan2024fosi} under a Poisson likelihood. Finally, profile likelihood analysis \cite{murphy2000profile, maiwald2016driving} ensures structural identifiability and yields calibrated confidence intervals for the inferred coupling parameter $\hat{\lambda}$.

To ensure the reliability of the proposed methodology, we first validate the inference pipeline against alternative optimization strategies and parametric variations. We then deploy the framework across three distinct empirical systems spanning biological, technological, and mobility domains: the whole-brain connectome of \textit{Drosophila melanogaster} \cite{dorkenwald2024neuronal}, the continental European power grid under progressive topological degradation \cite{strohmeier2021crowdsourced}, and the global air transportation network throughout the COVID-19 pandemic \cite{sun2020did}. In each setting, disentangling metric constraints from latent affinities provides an effective quantitative lens to probe structural organization and physical vulnerability.

\section{Methods}

\subsection{Spatio-Functional Model and Inference Pipeline}
\label{sec:spatio_functional_inference}

The spatio-functional inference framework operates as a self-contained, deterministic pipeline. Let $G = (V, E, W^{\text{obs}})$ be a weighted, undirected network of $N = |V|$ nodes. Each node $i$ possesses an observed physical coordinate vector $x_i \in \mathbb{R}^{d_{\text{phys}}}$ and an unobserved latent functional coordinate $s_i \in \mathbb{R}^{d_s}$, arranged as rows of matrices $X \in \mathbb{R}^{N \times d_{\text{phys}}}$ and $S \in \mathbb{R}^{N \times d_s}$, respectively. With physical pairwise distances denoted by $D^{\text{phys}}_{ij} = \|x_i - x_j\|_2$ and latent functional distances by $D^{\text{lat}}_{ij} = \|s_i - s_j\|_2$, connection affinities are determined by two distance-decay kernels:
\begin{equation}
K^{\text{geo}}_{ij} = \exp\left(-\frac{D^{\text{phys}}_{ij}}{r}\right), \quad K^{\text{func}}_{ij} = \exp\left(-\frac{D^{\text{lat}}_{ij}}{\sigma_0}\right),
\end{equation}
where $r > 0$ defines the characteristic physical interaction scale, and $\sigma_0$ fixes the gauge unit of the latent geometry. The raw expected connectivity kernel superimposes both regimes through the continuous coupling parameter $\lambda \in [0, 1]$:
\begin{equation}
W^{\text{raw}}_{ij} = (1 - \lambda) K^{\text{geo}}_{ij} + \lambda K^{\text{func}}_{ij}, \quad \forall i \neq j.
\end{equation}

Under this formulation, $\lambda$ acts as a conditional attribution parameter. By anchoring the null connectivity model to physical metric decay, the latent kernel is tasked with capturing non-metric functional routing that cannot be accounted for by Euclidean proximity alone. In topologies where functional communities are partially aligned with physical clustering, the metric kernel conservatively absorbs shared spatial variance, ensuring that non-zero estimates of $\lambda$ strictly reflect functional affinities that override geographic wiring constraints.

To decouple global parameter estimation from local density variations induced by high-degree hubs, raw affinities are symmetrically normalized into a connection propensity matrix $\widetilde{W}$:
\begin{equation}
\widetilde{W}_{ij} = \frac{1}{2} \left( \frac{W^{\text{raw}}_{ij}}{\sum_{k=1}^N W^{\text{raw}}_{ik}} + \frac{W^{\text{raw}}_{ij}}{\sum_{k=1}^N W^{\text{raw}}_{jk}} \right).
\end{equation}
This formulation enforces symmetry ($\widetilde{W}_{ij} = \widetilde{W}_{ji}$) while satisfying the total mass identity $\sum_{i,j} \widetilde{W}_{ij} = N$, ensuring an average nodal propensity of $\sum_{j=1}^N \widetilde{W}_{ij} \approx 1$ (Appendix~\ref{app:mass_conservation}). Expected Poisson link intensities are subsequently scaled by the empirical mean nodal strength $\langle s \rangle = \frac{1}{N} \sum_{i,j} W^{\text{obs}}_{ij}$:
\begin{equation}
W^{\text{pred}}_{ij} = \langle s \rangle \, \widetilde{W}_{ij},
\end{equation}
guaranteeing the exact conservation of total network weight, such that $\sum_{i,j} W^{\text{pred}}_{ij} = \sum_{i,j} W^{\text{obs}}_{ij}$.

Assuming discrete edge counts follow independent Poisson distributions conditioned on latent intensities, $W^{\text{obs}}_{ij} \sim \text{Poisson}(W^{\text{pred}}_{ij})$, the model parameters $\theta = \{\lambda, r, S\}$ are inferred by minimizing the regularized negative log-likelihood over unique pairs $i < j$:
\begin{equation}
\mathcal{L}(\theta) = \sum_{i < j} \left( W^{\text{pred}}_{ij} - W^{\text{obs}}_{ij} \ln W^{\text{pred}}_{ij} \right) + \frac{\gamma}{2} \left(\ln r - \ln r_0\right)^2.
\end{equation}
Under the Poisson pseudo-maximum likelihood (PPML) framework~\cite{gourieroux1984pseudo}, this objective yields statistically consistent parameter estimates whenever the conditional mean is correctly specified ($\mathbb{E}[W^{\text{obs}}_{ij}] = W^{\text{pred}}_{ij}$), extending the formulation to non-negative continuous weights (such as transmission capacities) while naturally accommodating structural zeros. The quadratic penalty introduces a weakly informative log-prior centered at the empirical nearest-neighbor baseline $r_0 = \frac{2}{N} \sum_{i=1}^N \min_{j \neq i} D^{\text{phys}}_{ij}$. Setting $\gamma = 0.25$, corresponding to a standard deviation of $\sigma_{\ln r} = 2.0$, prevents numerical divergence in early optimization phases without constraining the asymptotic data-driven estimate $\hat{r}$.

To navigate the non-convex optimization landscape and resolve rotational gauge ambiguities, initial functional positions $S^{(0)} \in \mathbb{R}^{N \times d_s}$ are computed via deterministic spectral decomposition of the connection residuals unexplained by geography. A spatial reference matrix $W^{\text{geo}}_0$ is constructed from $K^{\text{geo}}_0 = \exp(-D^{\text{phys}} / r_0)$ under balanced normalization. Residual intensities $R_{\text{init}} = W^{\text{obs}} - W^{\text{geo}}_0$ isolate pairs whose connectivity exceeds metric decay. Invoking the Eckart--Young--Mirsky theorem \cite{golub1987generalization}, the optimal rank-$d_s$ Frobenius-norm approximation of the thresholded positive residual matrix $R^+ = \max(0, R_{\text{init}})$ is obtained via a truncated Singular Value Decomposition (SVD), $R^+ \approx U_{d_s} \Sigma_{d_s} U_{d_s}^T$, yielding initial latent coordinates:
\begin{equation}
S^{(0)} = U_{d_s} \Sigma_{d_s}^{1/2},
\end{equation}
where $U_{d_s} \in \mathbb{R}^{N \times d_s}$ contains the leading singular vectors and $\Sigma_{d_s} \in \mathbb{R}^{d_s \times d_s}$ is the diagonal matrix of dominant singular values (coinciding with the dominant eigenpairs of $R^+$ by matrix symmetry). Each column of $S^{(0)}$ is standardized to zero mean and unit variance, matching the gauge scale $\sigma_0$ and ensuring well-conditioned initial gradients.

Optimization proceeds deterministically through a two-stage hybrid protocol \cite{sivan2024fosi}. An initial warm-up phase of 40 iterations with the Adam optimizer ($\eta_{\text{Adam}} = 0.04$) moves coordinates away from the spectral boundary into the dominant basin of attraction \cite{kingma2014adam}. This is followed by 40 iterations of the quasi-Newton L-BFGS algorithm \cite{liu1989limited}, which uses accumulated gradient histories to approximate the inverse Hessian and achieve rapid local convergence. Parameter uncertainty on $\hat{\lambda}$ is subsequently quantified via profile likelihood analysis over a perturbation grid $\lambda_{\pm} = \hat{\lambda} \pm \Delta$ with $\Delta = 0.04$:
\begin{equation}
\mathcal{L}_{\text{prof}}(\lambda) = \min_{r, S} \mathcal{L}(r, S \mid \lambda).
\end{equation}
The empirical Fisher curvature $d^2_{\text{NLL}} = [\mathcal{L}_{\text{prof}}(\hat{\lambda} + \Delta) - 2\mathcal{L}_{\text{prof}}(\hat{\lambda}) + \mathcal{L}_{\text{prof}}(\hat{\lambda} - \Delta)] / \Delta^2$ provides the asymptotic standard error $\sigma_{\lambda} = (d^2_{\text{NLL}})^{-1/2}$, yielding calibrated $95\%$ confidence intervals:
\begin{equation}
\text{CI}_{95\%}(\lambda) = \left[ \hat{\lambda} - 1.96 \sigma_{\lambda}, \; \hat{\lambda} + 1.96 \sigma_{\lambda} \right].
\end{equation}

\subsection{Synthetic Validation and Hyperparameter Robustness}

To assess reconstruction fidelity and verify the operational stability of the pipeline under controlled ground-truth conditions, we benchmarked the framework on synthetic networks with known spatial and functional coordinates. Graphs were synthesized across varying network sizes by jointly sampling true coupling parameters, characteristic physical interaction ranges, and latent scales, with latent positions generated via Gaussian Mixture Models to reproduce modular cluster structures (Appendix~\ref{app:benchmark_generation}).

Because the inferred latent space is invariant under global isometric transformations, including rigid rotations and reflections, reconstructed configurations are aligned to ground-truth geometries via Orthogonal Procrustes analysis to evaluate structural recovery (Appendix~\ref{app:procrustes_alignment} for the algebraic derivation).

Using this benchmark, systematic sensitivity analyses across Adam learning rates, exploratory step budgets (Appendix~\ref{app:adam_validation}), and quasi-Newton iteration limits (Appendix~\ref{app:quasi_newton_validation}) demonstrate good across all parameters. 

To evaluate identifiability beyond idealized generative conditions, we perturbed connection propensities with uniform maximum-entropy topological noise \cite{chen2013system} (Appendix~\ref{sec:perturbation_robustness}). The results demonstrate a clear resilience hierarchy: while the physical reach scale $r$ is sensitive to noise-induced spatial flattening, both the coupling parameter $\lambda$ and the latent geometry degrade due to spectral filtering. Finally, complementary stress tests confirm framework generalizability under heavy-tailed power-law distance decays \cite{barthelemy2003crossover, barthelemy2018transitions} (Appendix~\ref{sec:powerlaw_decay_analysis}) and delineate the trade-offs governing the choice of latent dimensionality $d_s$ \cite{loyal2025spike, gwee2025latent} (Appendix~\ref{sec:latent_dimension_analysis}).

\section{Results}
\noindent

\subsection{Continuous Parameter Identifiability and Phase-Space Landscapes}
\label{sec:results_identifiability_landscape}

Disentangling geographic decay from latent affinity is fundamentally challenging when diffuse spatial ranges or weak relational ties produce overlapping topological signatures. To establish these identifiability boundaries, we mapped the operational phase space across an ensemble of synthetic networks systematically sampled over the continuous plane $(\lambda_{\text{true}}, r_{\text{true}})$ (see Appendix~\ref{app:synthetic_benchmark} for the generative pipeline). We quantified inference fidelity across three complementary error metrics: the absolute mixing error $e_\lambda = \vert{}\hat{\lambda} - \lambda_{\text{true}}\vert{}$, the relative spatial reach error $e_r = \vert{}\hat{r} - r_{\text{true}}\vert{} / r_{\text{true}}$, and the latent metric space deficit $\mathcal{D}_{\text{lat}} = 1 - \rho_{\text{lat}} \in [0, 1]$, where $\rho_{\text{lat}}$ is the Pearson distance correlation between true and reconstructed latent coordinates. By doing so, we uncover distinct operational regimes and fundamental detection limits inherent to socio-spatial inverse problems. 

\begin{figure}[!htbp]
    \centering
    \includegraphics[width=\linewidth]{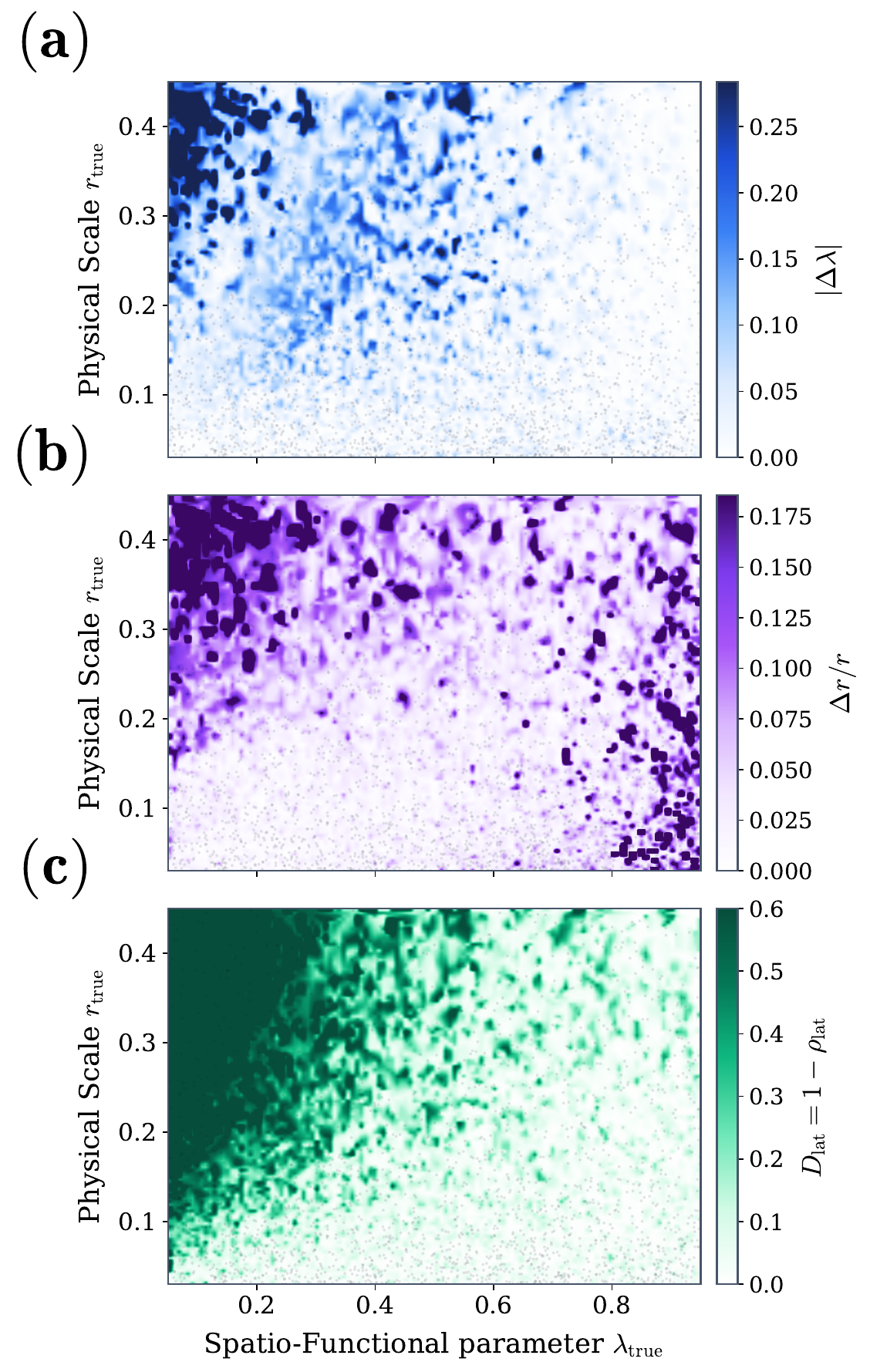}
    \caption{\textbf{Continuous phase-space identifiability landscapes across $N_{\text{nets}} = 10{,}000$ synthetic network realizations.} 
    Performance surfaces reconstructed via 2D Delaunay triangulation interpolation over the continuous generative plane $(\lambda_{\text{true}}, r_{\text{true}})$, evaluated across graph sizes $N \in [100, 2000]$. We sample true physical interaction scales $\ln r_{\text{true}} \sim \mathcal{U}(\ln 0.03, \ln 0.50)$, \textbf{(a)} Absolute mixing parameter error $e_\lambda = |\hat{\lambda} - \lambda_{\text{true}}|$.
    \textbf{(b)} Relative spatial reach error $e_r = |\hat{r} - r_{\text{true}}| / r_{\text{true}}$.
    \textbf{(c)} Latent metric space deficit $\mathcal{D}_{\text{lat}} = 1 - \rho_{\text{lat}}$.}
    \label{fig:identifiability_heatmaps}
\end{figure}

The macroscopic mixing parameter $\lambda$ demonstrates near-universal identifiability across the configuration space (Fig.~\ref{fig:identifiability_heatmaps}a). A bounded region of elevated parameter error ($e_\lambda \approx 0.15\text{--}0.25$) emerges strictly in the upper-left quadrant, where weak functional affinity ($\lambda_{\text{true}} \le 0.20$) coexists with macroscopic physical reach scales ($r_{\text{true}} \ge 0.35$). In this regime, because $r_{\text{true}}$ approaches the diameter of the embedding domain, the spatial decay kernel $\exp(-D^{\text{phys}}_{ij}/r)$ flattens toward an all-to-all connectivity baseline. Consequently, separating a weak latent functional signal from a diffuse, quasi-uniform geometric background becomes statistically ill-conditioned.

In contrast, inferring the physical reach scale $r$ exposes two distinct asymptotic challenges governed by the network's spatio-functional balance (Fig.~\ref{fig:identifiability_heatmaps}b). Spatial scale recovery is exceptionally accurate ($e_r \le 0.03$) throughout the local and intermediate regimes ($r_{\text{true}} \le 0.20$), where steep metric distance decay imposes sharp structural gradients. However, identifiability degrades under two limiting boundaries: the \textit{functionally dominated regime} ($\lambda_{\text{true}} \to 1.0$), where the vanishing multiplicative prefactor $(1 - \lambda)$ structurally suppresses the physical likelihood contribution, and the \textit{diffuse spatial limit} ($r_{\text{true}} > 0.35, \lambda_{\text{true}} \le 0.20$). In this latter domain, flat spatial likelihood gradients allow the Tikhonov penalty to bias $\hat{r}$ downward toward the local nearest-neighbor scale $r_0$. Crucially, this metric contraction causes the optimizer to misattribute long-range spatial dyads to the latent kernel, directly explaining the correlated elevation of $e_\lambda$ in the corresponding region of parameter space.

Finally, the latent metric deficit $\mathcal{D}_{\text{lat}} = 1 - \rho_{\text{lat}}$ (Fig.~\ref{fig:identifiability_heatmaps}c) reveals a diagonal detectability boundary governed by the competition between physical scale and functional coupling. When physical wiring is localized ($r_{\text{true}} \lesssim 0.15$), steep geographic decay gradients isolate residual affinities with high contrast, enabling faithful geometric recovery ($\mathcal{D}_{\text{lat}} \le 0.10$) even under weak mixing ($\lambda_{\text{true}} \approx 0.10$). Reconstruction degrades substantially only in the upper-left quadrant ($r_{\text{true}} \gtrsim 0.25$ and $\lambda_{\text{true}} \lesssim 0.30$), where a diffuse spatial background obscures subtle functional clustering, rendering latent coordinate inference statistically ill-conditioned. Outside this adverse regime, spectral decomposition of residual interactions reliably guides the optimizer into the globally isometric basin of attraction.

The fundamental identifiability and precision advantages of this generative formulation are further corroborated in Appendix~\ref{sec:gnn_vs_mle_benchmark}, where we benchmark the analytical optimizer against a supervised spatial Graph Neural Network (GNN) baseline tailored to jointly resolve physical and latent connectivity channels.

\subsection{Disentangling Wiring Economy and Functional Specificity in the Drosophila Connectome}
\label{sec:results_drosophila_connectome}

\begin{figure*}[!htbp]
    \centering
    \includegraphics[width=\textwidth]{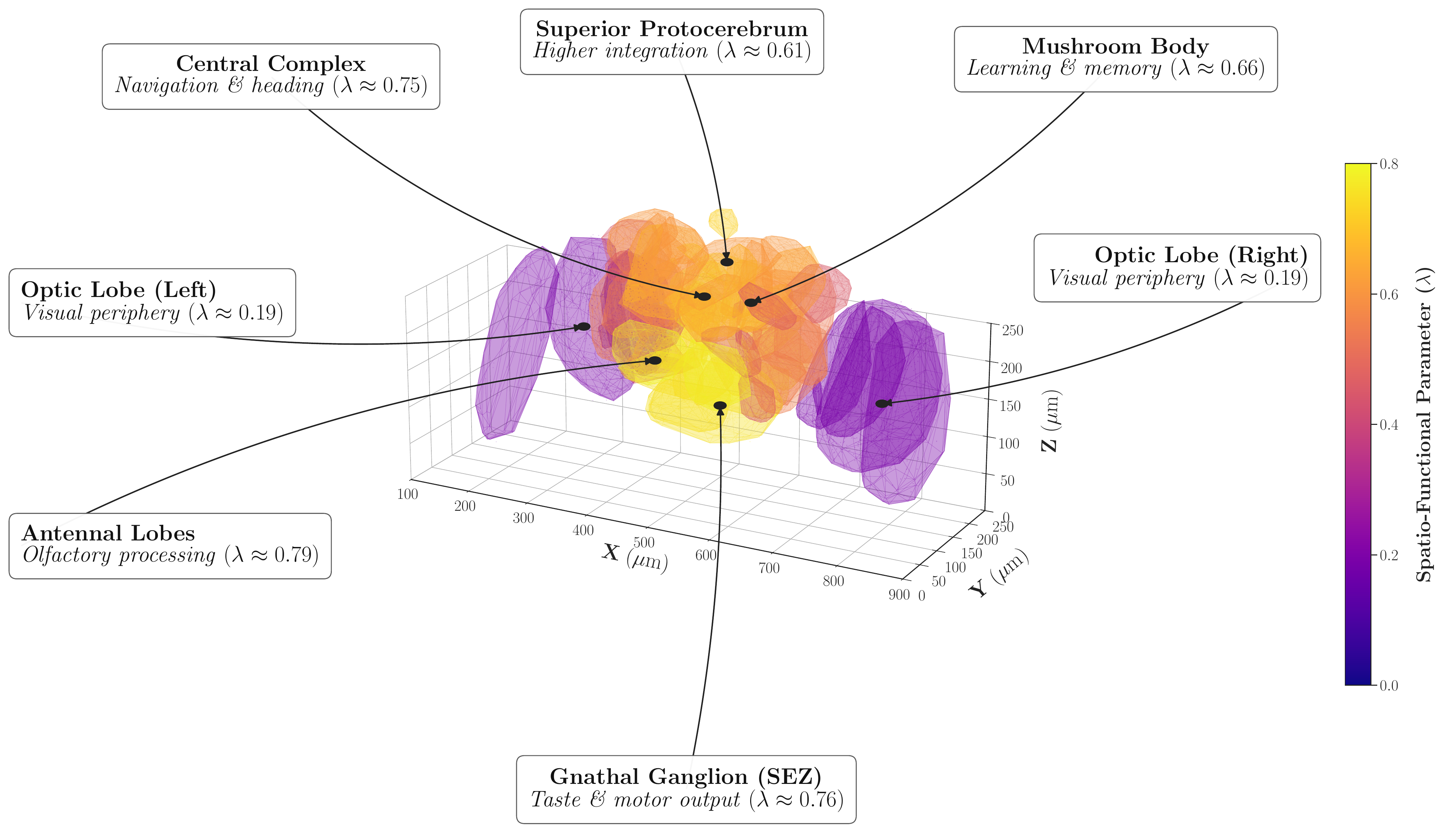}
    \caption{\textbf{Three-dimensional spatio-functional landscape of the \textit{Drosophila} whole-brain connectome.} 
    Reconstructed 3D layout of primary neuropil convex hulls ($X, Y, Z$ in $\mu\text{m}$), color-coded by the inferred spatio-functional coupling parameter $\hat{\lambda}$ via the plasma colormap. Callout labels highlight key functional systems across the topological spectrum, contrasting metric-dominated visual peripheries ($\lambda \approx 0.19$) with functionally specialized cognitive and integrative circuits, such as the Mushroom Body ($\lambda \approx 0.66$), Central Complex ($\lambda \approx 0.75$), and Antennal Lobes ($\lambda \approx 0.79$).}
    \label{fig:drosophila_connectome}
\end{figure*}

Having validated parameter identifiability on synthetic benchmarks and demonstrated computational scalability across large-scale graph regimes (Appendix~\ref{sec:system_size_scaling}), we examine whether the framework can uncover biological wiring principles in complex anatomical networks. To this end, we deployed the pipeline on the adult female \textit{Drosophila melanogaster} whole-brain connectome (FlyWire FAFB v783), comprising $139{,}255$ proofread neurons and over $80.2 \times 10^6$ synaptic contacts \cite{dorkenwald2024neuronal}. Parameterizing the latent functional coordinates in three dimensions to mirror the physical embedding space, we inferred the spatio-functional coupling parameter $\hat{\lambda} \in [0, 1]$ and characteristic physical reach scale $\hat{r}$ across its 13 major brain systems (Fig.~\ref{fig:drosophila_connectome} and Table~\ref{tab:drosophila_parameters}). In doing so, we test whether unsupervised statistical inference autonomously recapitulates the spectrum of physical wiring constraints and computational specializations established in the neurobiological literature.

\begin{table*}[!htbp]
\centering
\small
\caption{\textbf{Inferred spatio-functional coupling parameters across the 13 major brain systems of \textit{Drosophila melanogaster} (FAFB v783).} Parameters were estimated over $M = 50$ induced subgraph ensembles ($n = 600$ neurons per replica). Values are reported as mean $\pm$ standard deviation.}
\label{tab:drosophila_parameters}
\begin{tabular*}{\textwidth}{@{\extracolsep{\fill}} l c c c @{}}
\hline\hline\noalign{\smallskip}
\textbf{Brain Zone} & \textbf{Neurons ($N$)} & \textbf{Coupling} $\hat{\lambda} \pm \text{std}$ & \textbf{Spatial Scale} $\hat{r} \pm \text{std}$ [$\mu\mathrm{m}$] \\
\noalign{\smallskip}\hline\noalign{\smallskip}
\textbf{Antennal Lobes} & $3{,}237$ & $0.7869 \pm 0.0220$ & $61.25 \pm 13.91$ \\
\textbf{Inferior Protocerebrum \& Slopes} & $2{,}969$ & $0.7646 \pm 0.0252$ & $51.61 \pm 4.42$ \\
\textbf{Gnathal Ganglion (SEZ)} & $7{,}724$ & $0.7584 \pm 0.0386$ & $29.04 \pm 2.11$ \\
\textbf{Central Complex} & $2{,}794$ & $0.7472 \pm 0.0275$ & $34.28 \pm 3.00$ \\
\textbf{Other Periesophageal} & $1{,}019$ & $0.7279 \pm 0.0242$ & $43.96 \pm 2.09$ \\
\textbf{Mushroom Body} & $5{,}260$ & $0.6629 \pm 0.0770$ & $29.07 \pm 6.28$ \\
\textbf{Superior Protocerebrum} & $9{,}179$ & $0.6080 \pm 0.0297$ & $32.56 \pm 1.58$ \\
\textbf{Ventrolateral Protocerebrum} & $5{,}555$ & $0.5496 \pm 0.0356$ & $62.01 \pm 6.70$ \\
\textbf{Posterior Lateral Protocerebrum} & $1{,}465$ & $0.5326 \pm 0.0349$ & $36.92 \pm 2.93$ \\
\textbf{Anterior Optic Tubercle} & $903$ & $0.4787 \pm 0.0383$ & $79.87 \pm 6.23$ \\
\textbf{Lateral Complex \& Horn} & $3{,}174$ & $0.4556 \pm 0.0503$ & $43.35 \pm 3.51$ \\
\textbf{AMMC (Auditory/Mechanosensory)} & $692$ & $0.4300 \pm 0.0203$ & $33.95 \pm 1.14$ \\
\textbf{Optic Lobes} & $95{,}218$ & $0.1891 \pm 0.1143$ & $14.10 \pm 2.23$ \\
\noalign{\smallskip}\hline\hline
\end{tabular*}
\end{table*}

The inferred coupling parameters reveal that the insect brain does not adhere to a uniform organizational regime; rather, it spans a continuous spectrum ranging from strictly metric-driven sensory processors to abstract topological computing cores.

At the physical limit, the Optic Lobes are heavily constrained by three-dimensional metric space, exhibiting the lowest functional coupling ($\hat{\lambda} \approx 0.19$) and the steepest physical decay. This reflects the columnar, retinotopic architecture of the compound eye, where synaptic contacts within the lamina, medulla, and lobula occur almost exclusively between immediate physical neighbors to preserve visual-field topology with minimal metabolic wiring length \cite{rivera2011wiring, takemura2015synaptic}.

Intermediate regimes bridge sensory reception and premotor coordination. Neuropils such as the Superior Protocerebrum ($\hat{\lambda} \approx 0.61$) exhibit a balanced coexistence between spatial distance decay and specific pathway routing, mediating broad multi-sensory convergence and higher-level integration.

Conversely, higher-order integrative neuropils decouple from Euclidean constraints. The Mushroom Body displays elevated latent affinity ($\hat{\lambda} \approx 0.66$), consistent with the associative learning architecture of Kenyon cells and mushroom body output neurons, where synaptic plasticity overrides spatial proximity \cite{aso2014mushroom}. Higher still, the Central Complex operates in a predominantly topological regime ($\hat{\lambda} \approx 0.75$): its internal compass circuits within the ellipsoid body and fan-shaped body implement ring-attractor dynamics for head-direction representation and vector navigation, relying on non-local, algorithmically structured connectivity \cite{turner2020neuroanatomical}. Similarly, the Gnathal Ganglion ($\hat{\lambda} \approx 0.76$) reflects specialized sensorimotor circuits coordinating feeding behaviors and motor programs across dispersed descending pathways.

Finally, the Antennal Lobes display the highest functional specialization in the brain ($\hat{\lambda} \approx 0.79$). Olfactory receptor neurons project selectively onto homotypic projection neurons within dedicated glomeruli matching specific odorant receptor expressions, forming a chemo-affinity wiring network that completely subordinates geometric proximity to molecular identity \cite{hong2014genetic}.

These results reconcile a fundamental dichotomy in neuroanatomy, bridging metabolic wiring minimization, formalized by the Exponential Distance Rule (EDR)~\cite{ercsey2013predictive}, with functional topological routing. Rather than obeying a monolithic law, the Drosophila connectome demonstrates that metric economy governs only sensory peripheries ($\lambda \approx 0.19$), whereas associative, navigational, and chemotopic circuits progressively decouple from Euclidean space. The coupling parameter $\lambda$ thus provides a principled metric quantifying how computational demands override metabolic wiring costs across the nervous system~\cite{bullmore2012economy}.

\subsection{Regime Transitions and Structural Resilience in the European Power Grid}
\label{sec:results_power_grid_pruning}

\begin{figure*}[!htbp]
    \centering
    \includegraphics[width=\textwidth]{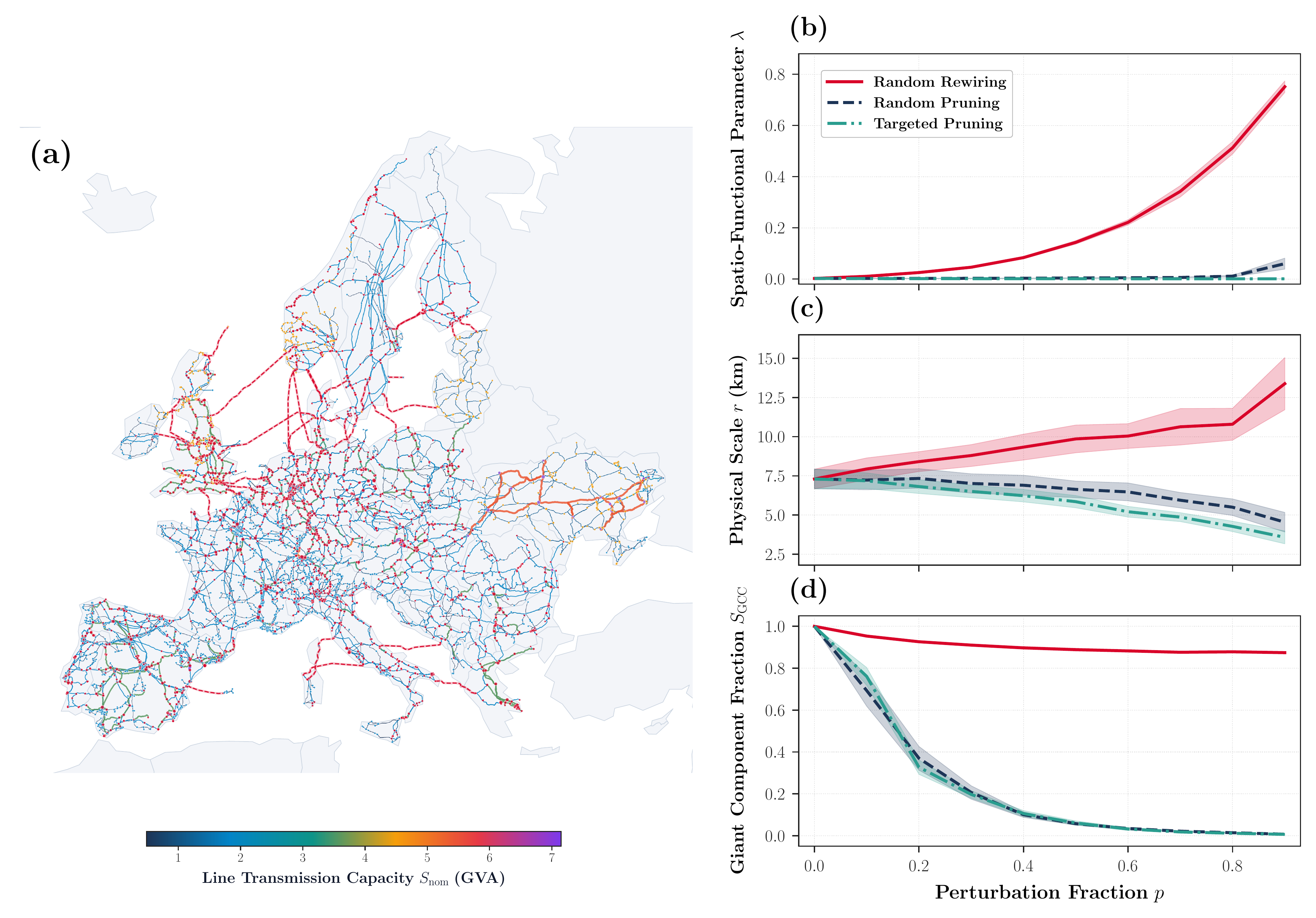}
    \caption{\textbf{Topological degradation and geometric regime transitions across the European high-voltage power grid.} 
    \textbf{(a)} Geographical layout of the continental transmission backbone ($220\text{--}750\,\text{kV}$ AC and physical HVDC links), with lines color-coded by nominal transmission capacity $S_{\text{nom}}$ in gigavolt-amperes (GVA).
    \textbf{(b)} Inferred spatio-functional coupling parameter $\hat{\lambda}$ as a function of edge perturbation fraction $p \in [0.0, 0.9]$.
    \textbf{(c)} Characteristic physical reach scale $\hat{r}$ (km).
    \textbf{(d)} Relative size of the giant connected component $S_{\text{GCC}}$. 
    Solid and dashed curves denote ensemble means over $M = 30$ regional subgraphs ($n = 600$ substations, base lines $E_{\text{base}} \approx 710$, initial mean degree $\langle k \rangle \approx 2.37$) with shaded 95\% confidence intervals; faint markers display individual subgraph realizations. Three edge-level protocols are evaluated: Random Rewiring (red solid line; lines reconnected to arbitrary substation pairs), Random Pruning (blue dashed line; uniform line removal), and Targeted Pruning (green dash-dotted line; line removal prioritized by capacity criticality $\mathrm{Score}(u,v) = s_u \cdot s_v$, where $s_u = \sum_w W_{uw}$ is the total substation capacity, selectively severing primary arteries connecting major electrical hubs).}
    \label{fig:europe_grid_benchmark}
\end{figure*}

To evaluate how the inferred parameters behave under structural degradation and topological transformations, we deployed the framework on the continental European high-voltage transmission grid (ENTSO-E / PyPSA; Fig.~\ref{fig:europe_grid_benchmark}a). Power grids represent quintessential physical networks: transmission lines adhere tightly to terrestrial geography, infrastructure investment constraints enforce near-planarity, and long-range bypasses skipping intermediate substations are energetically inefficient \cite{hartmann2026topology}. Subjecting regional grid subgraphs to systematic perturbations across a damage fraction $p \in [0.0, 0.9]$, we evaluated how the spatio-functional coupling $\hat{\lambda}$ and characteristic physical reach scale $\hat{r}$ evolve alongside the giant component fraction $S_{\text{GCC}}$ across three distinct damage protocols. While link pruning directly emulates plausible real-world contingencies and structural disruptions, random rewiring is introduced as a synthetic counterfactual control to probe the theoretical response of the pipeline when metric embedding is deliberately dismantled.

At the unperturbed baseline ($p = 0.0$), the infrastructure operates strictly in the metric domain: the inferred spatio-functional coupling collapses to $\hat{\lambda}_0 \approx 0.002$ with a characteristic physical interaction scale of $\hat{r}_0 \approx 7.3\,\text{km}$. This confirms that high-voltage power routing is governed almost exclusively by metric distance decay without latent non-spatial shortcuts.

Under progressive link removal, the model exhibits a fundamental structural invariance: across both \textit{Random Pruning} (emulating stochastic operational failures or localized environmental hazards) and \textit{Targeted Pruning} prioritized by transmission capacity (modeling targeted disruptions on critical transmission corridors) \cite{cuadra2015critical}, $\hat{\lambda}$ remains anchored near zero even at $90\%$ edge loss ($p = 0.9$; Fig.~\ref{fig:europe_grid_benchmark}b). Structural degradation thins the physical infrastructure without inducing spurious functional clustering. Concurrently, the physical reach scale $\hat{r}$ contracts monotonically from $7.3\,\text{km}$ down to $4.5\,\text{km}$ under random pruning and $3.6\,\text{km}$ under targeted pruning (Fig.~\ref{fig:europe_grid_benchmark}c), reflecting the progressive erosion of intermediate-range transmission corridors until only immediate geographic adjacencies survive.

Although non-local rewiring is physically and economically unfeasible in power systems engineering, this synthetic stress test triggers an explosive, superlinear transition in the coupling parameter (Fig.~\ref{fig:europe_grid_benchmark}b). Despite preserving the total link budget ($E \equiv \text{const}$, $\langle k \rangle \approx 2.37$), non-local shortcuts destroy the underlying geographic embedding, causing $\hat{\lambda}$ to surge from $\approx 0.002$ to $0.75$ at $p = 0.90$ as the latent space is recruited to absorb non-local interactions. Concurrently, $\hat{r}$ dilates toward $13.4\,\text{km}$ (Fig.~\ref{fig:europe_grid_benchmark}c) as the spatial kernel artificially stretches before yielding its explanatory power to the latent functional component.

Crucially, tracking the joint parameter trajectory $(\hat{\lambda}(p), \hat{r}(p))$ decouples physical integrity from classical percolation metrics (Fig.~\ref{fig:europe_grid_benchmark}d). While percolation analysis indicates catastrophic network fragmentation under pruning ($S_{\text{GCC}} \approx 0.33$ at $p = 0.20$ targeted, collapsing to $< 0.01$ at $p = 0.90$), it deems rewired grids highly resilient ($S_{\text{GCC}} \approx 0.87$ at $p = 0.90$) \cite{wandelt2022random} due to expander-like topological connectivity. This apparent robustness is fundamentally deceptive, as the rewired system completely abandons its spatial embedding and terrestrial constraints. Our framework thus clearly discriminates physical structural erosion ($\hat{\lambda} \approx 0$ with contracting $\hat{r}$) from topological delocalization (surging $\hat{\lambda}$), providing a principled diagnostic capable of disentangling geometric integrity from abstract topological connectivity in complex spatial networks.

\subsection{Dynamic Regime Shifts and Systemic Spatialization in Global Aviation During COVID-19}
\label{sec:results_aviation_covid}

Beyond static biological architectures and steady-state infrastructural degradation, we examine the diagnostic capacity of our framework to monitor dynamic regime shifts in non-stationary empirical systems. To this end, we tracked the macroscale reconfiguration of the global air transportation network across 48 consecutive months, spanning from January 2019 through December 2022. Using crowdsourced open-access ADS-B flight trajectory data curated by the OpenSky Network \cite{strohmeier2021crowdsourced}, we compiled monthly origin-destination flight volume matrices across commercial airports worldwide. 

\begin{figure}[!htbp]
    \centering
    \includegraphics[width=\columnwidth]{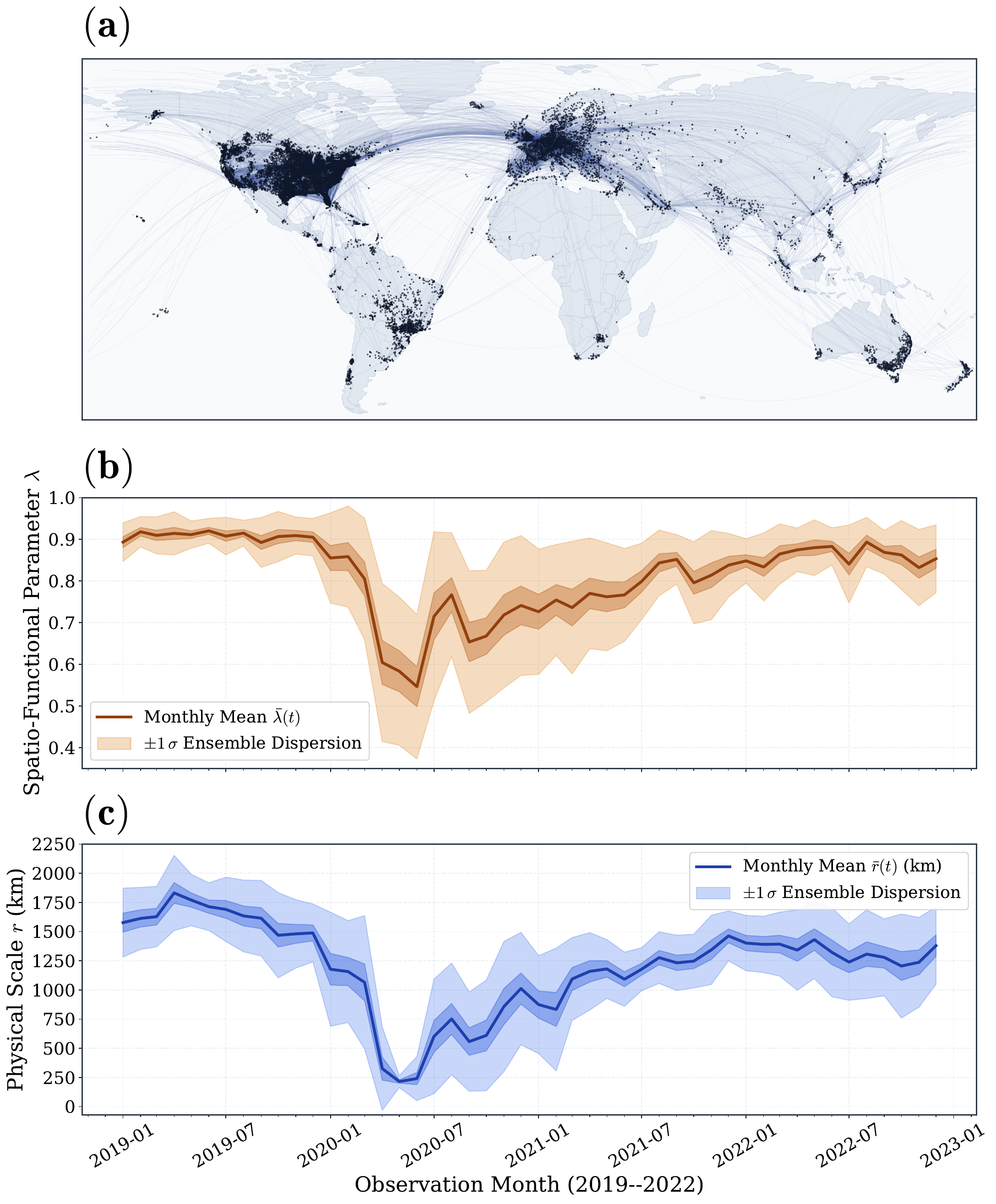}
    \caption{\textbf{Dynamic regime shifts and systemic spatialization of the global air transportation network during the COVID-19 pandemic (2019--2022).} 
    \textbf{(a)} Global airport connectivity network, where geodesic flight trajectories connect active commercial airport pairs with line thickness proportional to mean monthly flight volume across the entire 48-month observation window (2019--2022), delineating primary transoceanic corridors and continental hub clusters. 
    \textbf{(b)} Temporal trajectory of the inferred spatio-functional coupling parameter $\bar{\lambda}(t) \in [0, 1]$ (labeled as mixing parameter on the vertical axis).
    \textbf{(c)} Evolution of the characteristic physical reach scale $\bar{r}(t)$. Solid curves track monthly ensemble means across $M = 50$ induced subgraphs ($n = 600$ airports per replica); shaded envelopes denote the $\pm 1\sigma$ empirical ensemble dispersion across resamples (outer light band) and the 95\% confidence interval of the ensemble mean ($\pm 1.96\,\mathrm{SE}$, inner dark band).}
    \label{fig:aviation_covid_benchmark}
\end{figure}

Geographic coordinates were mapped onto three-dimensional chordal Euclidean distances on the terrestrial sphere, while the unobserved functional coordinates were parameterized in a flat two-dimensional Euclidean latent space \footnote{While terrestrial geography is intrinsically spherical, three-dimensional chordal distances furnish a smooth Euclidean metric free of geodesic branch cuts and antipodal singularities. In parallel, because socio-economic flight affinities, airline alliances, and global routing hierarchies do not possess intrinsic spherical closure or compact topological boundaries, parameterizing the latent coordinates in a flat Euclidean space $\mathbb{R}^{d_s}$ provides the minimal unconstrained model, preserving metric homogeneity between the physical and latent interaction kernels without imposing extraneous geometric priors.}. Tracking the joint parameter trajectory $(\hat{\lambda}(t), \hat{r}(t))$ across this standardized metric framework directly quantifies systemic spatialization throughout the pandemic lifecycle (Fig.~\ref{fig:aviation_covid_benchmark}).

Throughout the pre-pandemic baseline year (2019), global commercial aviation operated within an overwhelmingly topological and relational regime. The inferred spatio-functional coupling remained stably elevated at $\hat{\lambda} \approx 0.90\text{--}0.92$ (Fig.~\ref{fig:aviation_covid_benchmark}b), establishing that over $90\%$ of link propensities were driven by latent functional affinities, including airline alliances, multi-hub transfers, and economic centralities that deliberately override geographic distance decay \cite{guimera2005worldwide}. Concurrently, the physical reach scale hovered at macroscopic distances ($\hat{r} \approx 1{,}600\text{--}1{,}800\,\text{km}$; Fig.~\ref{fig:aviation_covid_benchmark}c), reflecting a well-connected transport system underpinned by transoceanic and intercontinental corridors.

The imposition of international border closures, travel restrictions, and quarantine mandates in the spring of 2020 triggered an acute structural dislocation, suppressing primary intercontinental routes (Fig.~\ref{fig:aviation_covid_benchmark}a) \cite{sun2020did}. Our framework registers this breakdown as a precipitous collapse in functional coupling, plunging to an unprecedented minimum of $\hat{\lambda} \approx 0.55$ in June 2020 (Fig.~\ref{fig:aviation_covid_benchmark}b). This shift marks an abrupt systemic spatialization of global aviation with long-range functional shortcuts severed.

This topological retraction is corroborated by the physical interaction scale $\hat{r}(t)$ (Fig.~\ref{fig:aviation_covid_benchmark}c), which contracted by 88\% relative to pre-pandemic baselines, bottoming out at $\hat{r} = 215.7\text{ km}$ in May 2020. Crucially, $\hat{r}(t)$ measures the effective reach of spatial decay rather than the average flight length. With intercontinental passenger travel largely suspended, connectivity collapsed into localized domestic routes governed by sharp distance decay, while the few surviving long-haul corridors were absorbed by the residual functional coupling ($\lambda \approx 0.55$).

Following the acute lockdown shock, tracking the parameter trajectories throughout 2021 and 2022 reveals strong structural hysteresis. While topological collapse occurred in a matter of weeks during early 2020, recovery was protracted, non-uniform, and geographically staggered. The characteristic physical reach scale $\hat{r}$ gradually rebounded toward $1{,}200\text{--}1{,}500\,\text{km}$, while the spatio-functional coupling $\hat{\lambda}$ steadily climbed back toward $0.85\text{--}0.88$ by late 2022. By remaining marginally below pre-pandemic baselines, these trajectories reflect the asynchronous reopening of bilateral flight corridors and persistent structural reorganization in global aviation networks \cite{sun2021impact}.

\section{Discussion}
\label{sec:discussions}

In this work, we formulated a generative statistical framework to solve the inverse problem of spatio-functional inference in weighted complex networks. By modeling edge weights as a dual-channel superposition of geographic distance decay and latent functional proximity, the framework decomposes observed topology into an explicit coupling parameter $\lambda \in [0, 1]$, an effective spatial interaction scale $r$, and a low-dimensional latent geometry $S$. To resolve the non-convex landscape and continuous rotational symmetries of the latent space, we introduced a deterministic pipeline pairing spectral residual initialization with hybrid first- and quasi-Newton optimization. Finally, profile likelihood analysis yields calibrated confidence intervals for parameter estimates, ensuring structural identifiability without relying on stochastic sampling.

This methodology offers a complementary bridge across spatial econometrics, statistical network modeling, and geometric deep learning. While gravity formulations traditionally account for non-spatial affinities through node-level fixed effects \cite{behrens2012dual}, spatial null models are primarily designed for statistical hypothesis testing rather than reconstructing unobserved latent coordinates \cite{sarzynska2016null}. On the other hand, while GNNs provide expressive representations for relational data, they operate largely as black-box architectures where spatial constraints and functional interactions cannot be easily separated into transparent physical parameters \cite{yehudai2021local}. In this landscape, our unsupervised approach provides an interpretable alternative: it operates directly on continuous weighted networks, explicitly isolates the relative weight of geographic decay via the scalar parameter $\lambda$, and maps unobserved connectivity into an inspectable latent geometry.

Applied across three empirical domains, this framework provided concrete insights into the interplay between spatial constraints and functional organization. In the whole-brain connectome of \textit{Drosophila melanogaster} \cite{dorkenwald2024neuronal}, unsupervised inference resolved an anatomical hierarchy transitioning from metric-dominated sensory peripheries to functionally specialized associative and integrative neuropils. In the continental European power grid, monitoring continuous trajectories in $(\hat{\lambda}, \hat{r})$ under topological degradation demonstrated that macroscopic topological indicators such as the giant component fraction and global efficiency \cite{albert2000error, latora2001efficient} can remain elevated while the underlying spatial integrity collapses. Finally, in the global aviation network, tracking parameter trajectories throughout the COVID-19 pandemic \cite{sun2020did} captured an abrupt systemic transition toward localized, distance-governed connectivity following the suspension of long-haul corridors. Together, these applications illustrate how disentangling metric constraints from latent affinities provides a unified perspective to investigate spatially embedded networks.

While hidden metric space frameworks and hyperbolic network geometries unify node popularity and topological similarity into a single curved manifold \cite{papadopoulos2012popularity, boguna2021network}, our dual-kernel formulation provides an explicit coupling parameter $\lambda \in [0, 1]$ that directly decouples physical wiring constraints from latent functional affinity. Looking forward, incorporating non-Euclidean latent geometries into this mixture, most notably hyperbolic spaces $\mathbb{H}^{d_s}$, could provide native geometric representations for functional channels exhibiting strict scale-free hierarchies and branching organizations \cite{krioukov2010hyperbolic}. In terms of temporal dynamics, formulating continuous-time tracking equations for $\lambda(t)$ in relational event streams would enable real-time anomaly detection and structural drift monitoring in streaming network data \cite{huang2022mutually}. On the empirical front, the framework can be extended to investigate multimodal urban mobility, such as separating localized pedestrian flows from high-speed underground transit in metropolitan areas \cite{ashraf2021impacts}. Furthermore, it provides a principled foundation for macroscale human connectomics, where mapping the balance between physical wiring economy and higher-order functional integration across neurodegenerative trajectories remains a central challenge \cite{sporns2011human}.

\section*{Data availability}
\noindent The whole-brain connectome dataset of the adult female \textit{Drosophila melanogaster} (FAFB v783) analyzed in this work is publicly available via the FlyWire Connectome Data Explorer (Codex) portal at \url{https://codex.flywire.ai}. Synaptic connectivity tables, 3D spatial coordinates of synaptic contacts, proofread neuron annotations, and neuropil mesh segmentations were obtained directly from the Codex public data releases and bulk downloads. The continental European high-voltage power grid dataset (ENTSO-E/PyPSA-Eur), comprising georeferenced substations, AC transmission lines, and physical HVDC links, is openly accessible via Zenodo at \url{https://doi.org/10.5281/zenodo.14144752}. The global air transportation dataset tracking flight operations and origin--destination connections throughout the COVID-19 pandemic, compiled from the OpenSky Network crowdsourced ADS-B receiver infrastructure, is openly accessible via Zenodo at \url{https://doi.org/10.5281/zenodo.7923702}. Corresponding airport geographic coordinates, ICAO identifiers, and terrestrial reference geometries were obtained from the open-access OurAirports public repository at \url{https://ourairports.com/data/}.

\section*{Code availability}
\noindent The main source code, PyTorch optimization routines, synthetic benchmark generators, and analysis pipelines developed in this study are openly accessible in the GitHub repository at \url{https://github.com/gabrielecerioli/Inverse_Inference_Weighted_Spatio_Functional_Networks}.

\section*{Authors' contributions}
\noindent GC implemented the computational framework, conducted all numerical simulations, and generated the figures. AC conceived the original idea, supervised the project, and wrote the initial draft of the manuscript. Both authors jointly contributed to the development of the methodological pipeline, analyzed the results, and critically reviewed and approved the final manuscript.

\appendix

\section{Propensity Matrix Normalization and Exact Mass Conservation}
\label{app:mass_conservation}

In the main text, raw affinities $W^{\text{raw}}_{ij}$ are normalized into the symmetric connection propensity matrix $\widetilde{W}_{ij}$ to decouple global parameter estimation from local hub density variations. Here, we present the complete algebraic derivation establishing that $\widetilde{W}$ strictly conserves total network mass, analyze the structural conditions governing local deviations from unity, and prove that the predicted intensity matrix $W^{\text{pred}}$ matches the empirical network volume.

Let $W^{\text{raw}} \in \mathbb{R}^{N \times N}$ denote the symmetric affinity matrix satisfying $W^{\text{raw}}_{ij} = W^{\text{raw}}_{ji}$. The unnormalized nodal strength is defined as $d_i = \sum_{k=1}^N W^{\text{raw}}_{ik}$. While the standard random-walk transition operator $P_{ij} = W^{\text{raw}}_{ij} / d_i$ satisfies strict row-stochasticity ($\sum_{j=1}^N P_{ij} = 1$), it breaks matrix symmetry whenever nodal degrees are heterogeneous ($P_{ij} \neq P_{ji}$). To preserve the undirected nature of the network, the connection propensity matrix $\widetilde{W}$ is constructed as the arithmetic mean of the row- and column-normalized operators:
\begin{equation}
\widetilde{W}_{ij} = \frac{1}{2} \left( \frac{W^{\text{raw}}_{ij}}{d_i} + \frac{W^{\text{raw}}_{ij}}{d_j} \right).
\end{equation}
By symmetry of the raw affinities, $\widetilde{W}_{ij} = \widetilde{W}_{ji}$. Summing over all pairs $(i, j)$ demonstrates the exact conservation of total mass:
\begin{align}
\sum_{i=1}^N \sum_{j=1}^N \widetilde{W}_{ij} &= \frac{1}{2} \sum_{i=1}^N \frac{1}{d_i} \left( \sum_{j=1}^N W^{\text{raw}}_{ij} \right) + \frac{1}{2} \sum_{j=1}^N \frac{1}{d_j} \left( \sum_{i=1}^N W^{\text{raw}}_{ij} \right) \notag \\
&= \frac{1}{2} \sum_{i=1}^N \frac{d_i}{d_i} + \frac{1}{2} \sum_{j=1}^N \frac{d_j}{d_j} = \frac{1}{2} N + \frac{1}{2} N = N.
\end{align}
This identity holds unconditionally across all values of the model parameters $\{\lambda, r, S\}$. Consequently, the network-wide average nodal propensity is strictly equal to unity:
\begin{equation}
\frac{1}{N} \sum_{i=1}^N \left( \sum_{j=1}^N \widetilde{W}_{ij} \right) = \frac{N}{N} = 1.
\end{equation}

Although the global mean is exactly 1, individual row sums $S_i = \sum_{j=1}^N \widetilde{W}_{ij}$ generally satisfy $S_i \approx 1$ rather than strict equality. Expanding the row sum for node $i$ yields:
\begin{equation}
\begin{aligned}
S_i = \sum_{j=1}^N \widetilde{W}_{ij} &= \frac{1}{2 d_i} \sum_{j=1}^N W^{\text{raw}}_{ij} + \frac{1}{2} \sum_{j=1}^N \frac{W^{\text{raw}}_{ij}}{d_j} \\
&= \frac{1}{2} + \frac{1}{2} \sum_{j=1}^N \frac{W^{\text{raw}}_{ij}}{d_j}.
\end{aligned}
\end{equation}
Achieving $S_i = 1$ strictly requires the second term to equal $1/2$, which corresponds to the condition $\sum_{j=1}^N (W^{\text{raw}}_{ij} / d_j) = 1$. This condition is satisfied identically in degree-regular networks where every node has uniform capacity, $d_j = d$ for all $j \in V$, reducing the summation to $\frac{1}{d} \sum_{j=1}^N W^{\text{raw}}_{ij} = \frac{d_i}{d} = 1$.

In heterogeneous topologies, departures from unity reflect structural disassortativity between hubs and peripheral nodes. For a peripheral node $i$ whose connectivity is concentrated almost entirely on a dominant hub $H$, we have $W^{\text{raw}}_{iH} \approx d_i$. Because the hub's total capacity is substantially larger ($d_H \gg d_i$), the corresponding ratio vanishes ($W^{\text{raw}}_{iH} / d_H \to 0$), driving the second term to zero and establishing a lower bound of $S_i \approx 1/2$. Conversely, for a central hub $H$ connected to numerous peripheral leaves $j$, each leaf directs the majority of its connection capacity toward the hub such that $W^{\text{raw}}_{Hj} / d_j \approx 1$. Summing across many such neighbors causes $\sum_j (W^{\text{raw}}_{Hj} / d_j) > 1$, resulting in $S_H > 1$. Enforcing strict doubly-stochastic scaling would require iterative procedures such as the Sinkhorn-Knopp algorithm, which would introduce considerable computational overhead and complicate gradient propagation within the optimization loop. The closed-form symmetrization used here provides an efficient first-order relaxation: it prevents high-degree hubs from dominating the Poisson likelihood while preserving exact total mass conservation.

Finally, scaling $\widetilde{W}$ by the empirical mean nodal strength $\langle s \rangle = \frac{1}{N} \sum_{i,j} W^{\text{obs}}_{ij}$ yields the predicted Poisson link intensity matrix:
\begin{equation}
W^{\text{pred}}_{ij} = \langle s \rangle \, \widetilde{W}_{ij}.
\end{equation}
Evaluating the sum across all node pairs confirms exact zero-th moment matching:
\begin{equation}
\sum_{i,j} W^{\text{pred}}_{ij} = \langle s \rangle \sum_{i,j} \widetilde{W}_{ij} = \left( \frac{1}{N} \sum_{i,j} W^{\text{obs}}_{ij} \right) N = \sum_{i,j} W^{\text{obs}}_{ij}.
\end{equation}
This property guarantees that the likelihood objective optimizes the spatial and topological distribution of connection weights without artificially inflating or deflating the overall network volume.

\section{Synthetic Network Benchmark and Procrustes Alignment}
\label{app:synthetic_benchmark}

\subsection{Synthetic Benchmark Generation Protocol}
\label{app:benchmark_generation}

To rigorously assess parameter identifiability, estimation consistency, and metric recovery fidelity in our dual inference framework, we developed an automated synthetic benchmark generating heterogeneous networks across a broad spectrum of topological configurations. Each synthetic instance is parameterized by graph size $N$, socio-spatial coupling $\lambda_{\text{true}}$, physical scale $r_{\text{true}}$, and latent scale $\sigma_{\text{true}}$. Network size is drawn uniformly from $N \sim \mathcal{U}\{100, 300\}$, while the mixing parameter is sampled as $\lambda_{\text{true}} \sim \mathcal{U}(0, 1)$, continuously interpolating between pure physical distance decay ($\lambda = 0$) and pure latent functional routing ($\lambda = 1$). The characteristic length scales are sampled log-uniformly over an order of magnitude:
\begin{equation}
    \ln r_{\text{true}}, \, \ln \sigma_{\text{true}} \sim \mathcal{U}\left(\ln 0.03, \, \ln 0.50\right).
\end{equation}

Node positions in physical space, $X = (x_1, \dots, x_N)^T \in \mathbb{R}^{N \times 2}$, are drawn uniformly on the unit square, $x_i \sim \mathcal{U}([0, 1]^2)$. To emulate realistic community clustering and homophily, ground-truth latent coordinates $S_{\text{true}} \in \mathbb{R}^{N \times d_s}$ (with $d_s = 2$) are generated from a Gaussian Mixture Model with $K \sim \mathcal{U}\{2, \dots, 8\}$ components:
\begin{equation}
    p(s) = \sum_{k=1}^K \pi_k \, \mathcal{N}\left(s \,\middle|\, \mu_k, \, \Sigma_k\right),
\end{equation}
where mixing weights follow $\boldsymbol{\pi} \sim \text{Dirichlet}(\mathbf{1}_K)$, centroids are sampled as $\mu_k \sim \mathcal{U}([0.1, 0.9]^{d_s})$, and covariance matrices are diagonal with elements $\sigma_{k,d}^2 \sim \mathcal{U}(0.01, 0.05)$. The resulting matrix $S_{\text{true}}$ is min-max normalized coordinate-wise to reside strictly within $[0, 1]^{d_s}$, matching the bounded support of the physical domain.

Given physical and latent pairwise Euclidean distance matrices, $D^{\text{phys}}_{ij} = \|x_i - x_j\|_2$ and $D^{\text{lat}}_{ij} = \|s_i - s_j\|_2$, raw affinity weights $W^{\text{raw}} \in \mathbb{R}^{N \times N}_{\ge 0}$ are computed via the dual-kernel mixture:
\begin{equation}
    W^{\text{raw}}_{ij} = (1 - \lambda_{\text{true}}) \exp\left(-\frac{D^{\text{phys}}_{ij}}{r_{\text{true}}}\right) + \lambda_{\text{true}} \exp\left(-\frac{D^{\text{lat}}_{ij}}{\sigma_{\text{true}}}\right),
\end{equation}
for all distinct pairs $i \neq j$, with vanishing diagonal entries ($W^{\text{raw}}_{ii} = 0$). To account for degree heterogeneity while preserving symmetry, raw affinities are mapped to a doubly-balanced connection propensity matrix:
\begin{equation}
    \widetilde{W}_{ij} = \frac{1}{2} \left( 
    \frac{W^{\text{raw}}_{ij}}{\sum_{k=1}^N W^{\text{raw}}_{ik}} 
    + \frac{W^{\text{raw}}_{ij}}{\sum_{k=1}^N W^{\text{raw}}_{jk}} 
    \right).
\end{equation}
Setting the target average nodal strength to scale linearly with network size ($\langle s \rangle = N$) yields the expected intensity matrix $W^{\text{pred}}_{ij} = N \, \widetilde{W}_{ij}$. Undirected observed edge counts $W^{\text{obs}}_{ij}$ are then sampled independently for each pair $i < j$ from a conditional Poisson observation process:
\begin{equation}
    W^{\text{obs}}_{ij} \sim \text{Poisson}\left(W^{\text{pred}}_{ij}\right), \qquad W^{\text{obs}}_{ji} = W^{\text{obs}}_{ij},
\end{equation}
with $W^{\text{obs}}_{ii} = 0$. This formulation guarantees that unobserved links ($W^{\text{obs}}_{ij} = 0$) emerge naturally whenever expected propensities are small ($W^{\text{pred}}_{ij} \ll 1$), reproducing empirical network sparsity alongside counting noise.

\subsection{Orthogonal Procrustes Alignment and Metric Fidelity}
\label{app:procrustes_alignment}

Because the latent kernel depends exclusively on Euclidean distances $\|s_i - s_j\|_2$, the model likelihood is invariant under the Euclidean isometry group $\text{ISO}(d_s) = \mathbb{R}^{d_s} \rtimes \mathcal{O}(d_s)$, which includes arbitrary global translations ($S \mapsto S + \mathbf{1}_N t^T$) and orthogonal transformations ($S \mapsto S R$, with $R^T R = I_{d_s}$). Furthermore, fixing the gauge scale $\sigma_{\text{gauge}} = 1.0$ introduces a global scaling degree of freedom relative to unconstrained ground-truth coordinates. Consequently, directly comparing inferred configurations $\hat{S} \in \mathbb{R}^{N \times d_s}$ against $S_{\text{true}}$ requires resolving these isometric gauge symmetries.

Translational ambiguity is eliminated by centering both coordinate matrices to zero mean via the centering projector $H = I_N - \frac{1}{N}\mathbf{1}_N\mathbf{1}_N^T$, yielding $\bar{S}_{\text{true}} = H S_{\text{true}}$ and $\bar{\hat{S}} = H \hat{S}$. The optimal rigid alignment between the centered configurations is formulated as the scaled Orthogonal Procrustes problem:
\begin{equation}
    \min_{R \in \mathcal{O}(d_s), \, \delta > 0} \left\| \bar{S}_{\text{true}} - \delta \bar{\hat{S}} R \right\|_F^2, \quad \text{subject to} \quad R^T R = I_{d_s}.
    \label{eq:procrustes_obj}
\end{equation}
Expanding the Frobenius norm via the trace identity $\|A\|_F^2 = \text{Tr}(A^T A)$ gives:
\begin{equation}
    \left\| \bar{S}_{\text{true}} - \delta \bar{\hat{S}} R \right\|_F^2 = \left\|\bar{S}_{\text{true}}\right\|_F^2 + \delta^2 \left\|\bar{\hat{S}}\right\|_F^2 - 2\delta \text{Tr}\left(R^T M\right),
\end{equation}
where $M \equiv \bar{\hat{S}}^T \bar{S}_{\text{true}} \in \mathbb{R}^{d_s \times d_s}$ is the cross-covariance dispersion matrix. For any fixed scaling $\delta > 0$, minimizing residual error corresponds to maximizing $\text{Tr}(R^T M)$. 

Let $M = U \Sigma V^T$ be the singular value decomposition of $M$, where $U, V \in \mathcal{O}(d_s)$ and $\Sigma = \text{diag}(\sigma_1, \dots, \sigma_{d_s})$ contains non-negative singular values $\sigma_1 \ge \dots \ge \sigma_{d_s} \ge 0$. Using the cyclic property of the trace:
\begin{equation}
\begin{split}
    \text{Tr}\left(R^T M\right) &= \text{Tr}\left(R^T U \Sigma V^T\right) \\
    &= \text{Tr}\left(V^T R^T U \Sigma\right) = \sum_{k=1}^{d_s} Z_{kk} \sigma_k,
\end{split}
\end{equation}
where $Z \equiv V^T R^T U \in \mathcal{O}(d_s)$. Because $Z$ is orthogonal, its diagonal entries satisfy $|Z_{kk}| \le 1$. Given $\sigma_k \ge 0$, the sum is maximized if and only if $Z_{kk} = 1$ for all $k$, which implies $Z = I_{d_s}$. Setting $V^T R^T U = I_{d_s}$ directly determines the optimal rotation-reflection matrix:
\begin{equation}
    R^* = U V^T.
    \label{eq:optimal_rotation}
\end{equation}
Differentiating the residual error with respect to $\delta$ and setting the derivative to zero yields the optimal isotropic scale factor:
\begin{equation}
    \delta^* = \frac{\text{Tr}\left({R^*}^T M\right)}{\left\|\bar{\hat{S}}\right\|_F^2} = \frac{\sum_{k=1}^{d_s} \sigma_k}{\left\|\bar{\hat{S}}\right\|_F^2}.
\end{equation}
The optimal rigidly aligned latent configuration is therefore $S_{\text{aligned}} = \delta^* \bar{\hat{S}} R^*$.

Complementary to coordinate alignment, gauge-invariant structural recovery is quantified by evaluating the Pearson correlation coefficient across all $N(N-1)/2$ unique pairwise Euclidean distances in the latent space:
\begin{equation}
    \rho_{\text{lat}} \equiv \frac{\sum_{i < j} \Delta D^{\text{lat}}_{ij}(\hat{S}) \, \Delta D^{\text{lat}}_{ij}(S_{\text{true}})}{
    \sqrt{\sum_{i < j} \left(\Delta D^{\text{lat}}_{ij}(\hat{S})\right)^2} 
    \sqrt{\sum_{i < j} \left(\Delta D^{\text{lat}}_{ij}(S_{\text{true}})\right)^2}
    },
\end{equation}
where $\Delta D^{\text{lat}}_{ij}(A) \equiv D^{\text{lat}}_{ij}(A) - \bar{D}^{\text{lat}}(A)$ denotes deviations from the mean pairwise distance. Because $\rho_{\text{lat}}$ is strictly invariant under translations, rotations, reflections, and isotropic scalings, values close to unity ($\rho_{\text{lat}} \to 1.0$) confirm that the optimization pipeline recovers true latent relational geometries without metric distortion.

\section{Hyperparameter Robustness and Optimization Calibration}
\label{app:hyperparameter_robustness}

\subsection{Adam Learning Rate and Exploratory Steps Calibration}
\label{app:adam_validation}

To systematically calibrate the first-order exploratory phase before transitioning to quasi-Newton deterministic refinement, we conducted a two-dimensional grid search evaluating the joint sensitivity across the Adam learning rate $\eta_{\text{Adam}} \in [0.01, 0.32]$ and the warm-up iteration budget $T_{\text{Adam}} \in [10, 60]$. Performance was jointly quantified through the parameter estimation error $|\hat{\lambda} - \lambda_{\text{true}}|$ and the latent structure error $1 - \rho_{\text{lat}}$ across synthetic benchmark ensembles.

\begin{figure}[!htbp]
    \centering
    \includegraphics[width=\linewidth]{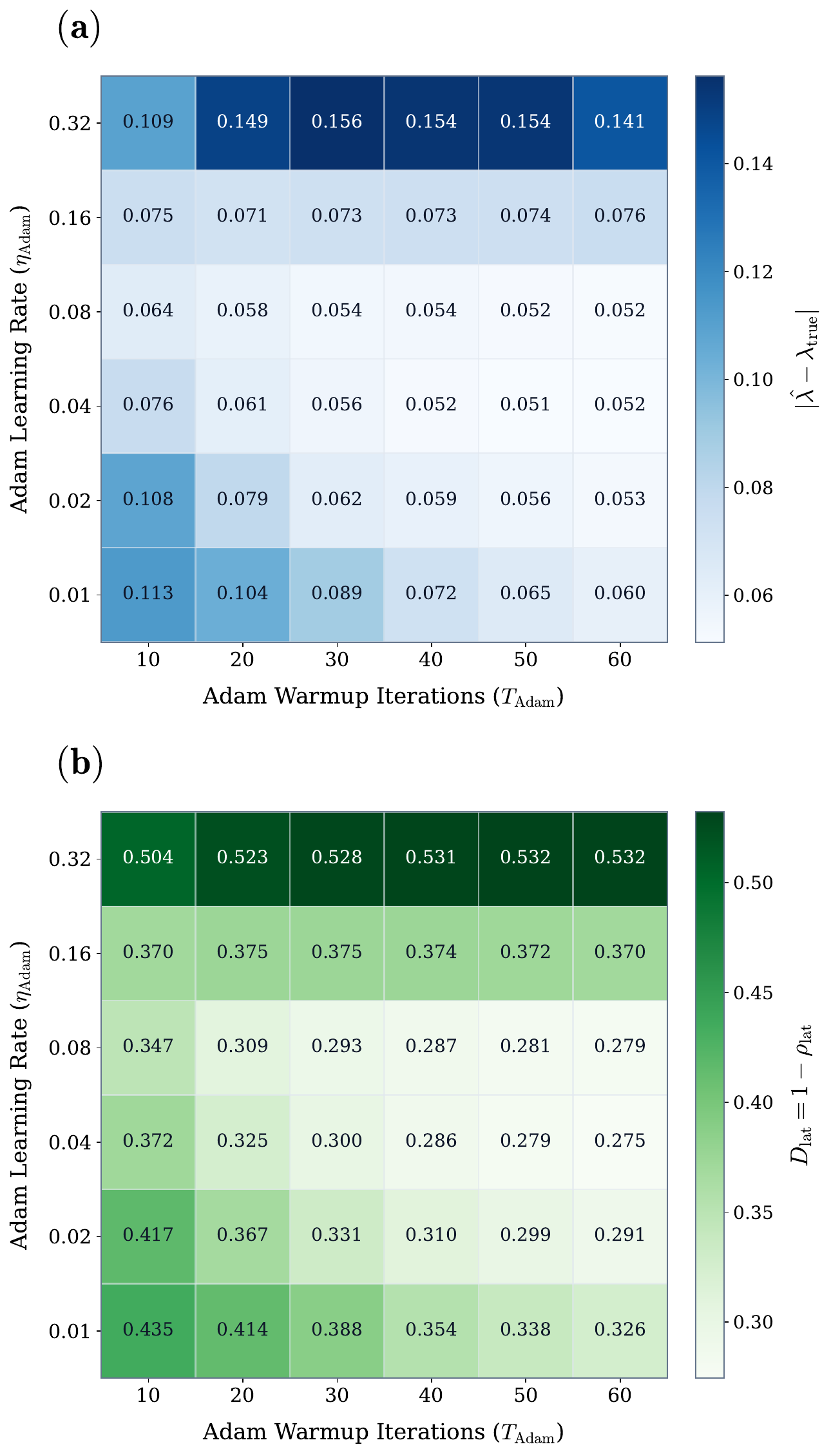}
    \caption{\textbf{Joint calibration of the first-order exploratory optimization phase.} 
    Performance heatmaps evaluated across Adam learning rates $\eta_{\text{Adam}} \in \{0.01, 0.02, 0.04, 0.08, 0.16, 0.32\}$ and warm-up step budgets $T_{\text{Adam}} \in \{10, 20, 30, 40, 50, 60\}$, followed by a fixed refinement budget of $T_{\text{L-BFGS}} = 40$ quasi-Newton iterations. Values represent averages over 100 independent synthetic network realizations per hyperparameter pair. \textbf{(a)} Mean parameter estimation error $|\hat{\lambda} - \lambda_{\text{true}}|$. \textbf{(b)} Mean latent structure error $(1 - \rho_{\text{lat}})$.}
    \label{fig:adam_calibration_heatmap}
\end{figure}

The empirical error landscapes delineate three distinct optimization regimes (Fig.~\ref{fig:adam_calibration_heatmap}). Conservative step sizes ($\eta_{\text{Adam}} \le 0.02$) suffer from under-exploration, yielding elevated parameter bias ($|\hat{\lambda} - \lambda_{\text{true}}| > 0.10$) and poor latent recovery ($1 - \rho_{\text{lat}} \ge 0.417$) at low iteration budgets ($T_{\text{Adam}} \le 20$). Conversely, excessively aggressive updates ($\eta_{\text{Adam}} \ge 0.16$) destabilize the non-convex trajectory; at $\eta_{\text{Adam}} = 0.32$, step overshoot ejects the optimizer from the attraction basin, inflating parameter error $|\hat{\lambda} - \lambda_{\text{true}}|$ and severely degrading latent alignment $1 - \rho_{\text{lat}}$.

The optimal basin of attraction spans $\eta_{\text{Adam}} \in [0.04, 0.08]$, where the exploratory momentum successfully relaxes the spectral initialization toward the global minimum. Within this corridor, parameter error drops to a stable plateau ($|\hat{\lambda} - \lambda_{\text{true}}| \approx 0.05$) and latent structure error reaches its minimum ($1 - \rho_{\text{lat}} \approx 0.28$) once $T_{\text{Adam}} \ge 40$.

Consequently, setting the default warm-up schedule to $\eta_{\text{Adam}} = 0.04$ with an exploration budget of $T_{\text{Adam}} = 40$ ensures minimal parameter bias and high latent metric recovery across all network topologies.

\subsection{Quasi-Newton Polish and Iteration Budget Calibration}
\label{app:quasi_newton_validation}

Following the first-order exploratory phase ($\eta_{\text{Adam}} = 0.04$, $T_{\text{Adam}} = 40$), the second stage employs the quasi-Newton L-BFGS routine to achieve superlinear convergence toward the deterministic optimum. To identify the optimal iteration budget and prevent overfitting to discrete Poisson fluctuations, we evaluated convergence across $T_{\text{L-BFGS}} \in [10, 640]$ over 100 independent synthetic network realizations.

\begin{figure}[!htbp]
    \centering
    \includegraphics[width=\linewidth]{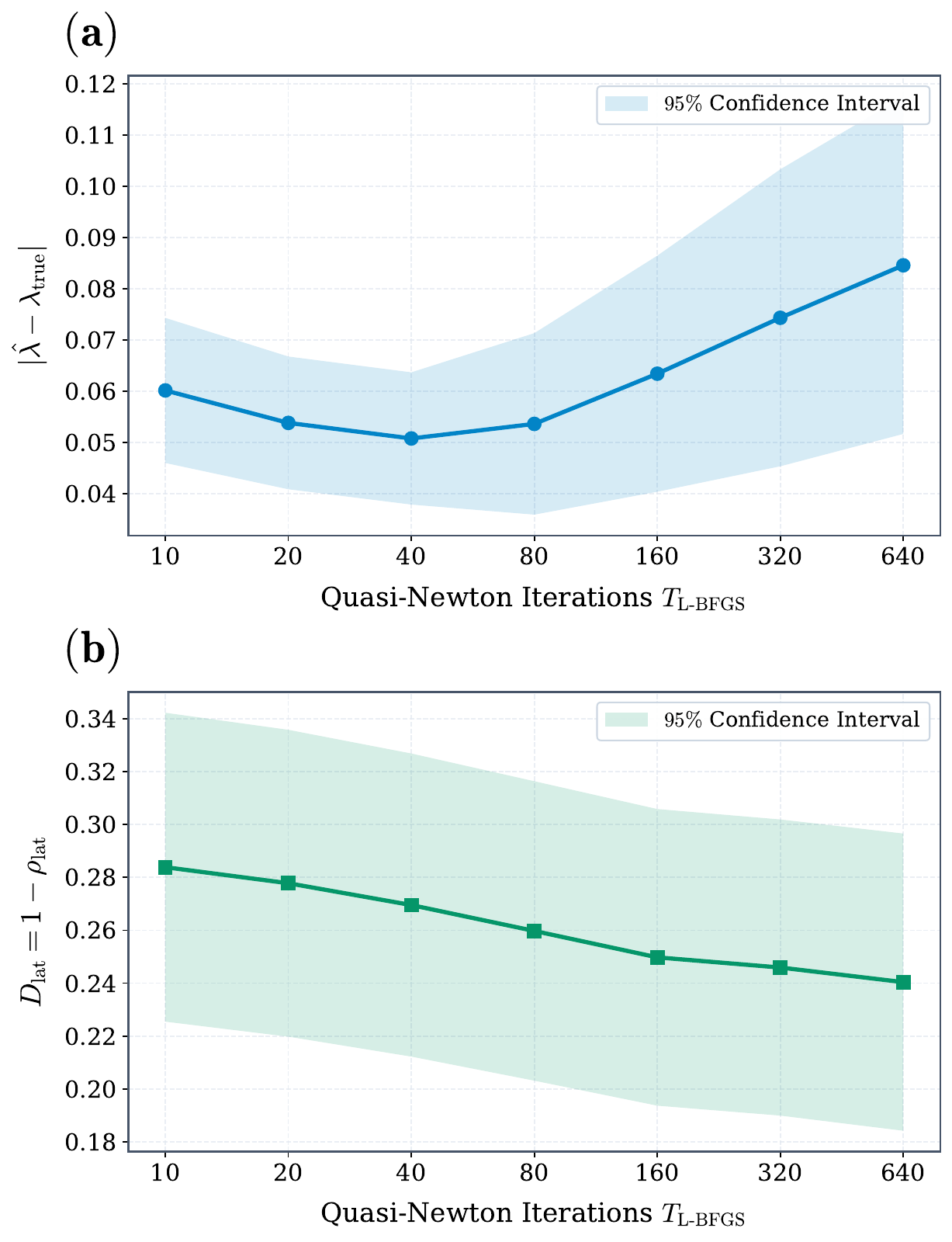}
    \caption{\textbf{Sensitivity analysis and convergence calibration of the L-BFGS phase.} Metrics evaluated across quasi-Newton iteration budgets $T_{\text{L-BFGS}} \in \{10, 20, 40, 80, 160, 320, 640\}$ on a $\log_2$ scale initialized after $T_{\text{Adam}} = 40$ warm-up steps and $\eta_{\text{Adam}}=0.04$. Shaded bands denote empirical 95\% confidence intervals across 100 synthetic network realizations. \textbf{(a)} Mean parameter estimation error $|\hat{\lambda} - \lambda_{\text{true}}|$. \textbf{(b)} Latent metric space deficit $\mathcal{D}_{\text{lat}} = 1 - \rho_{\text{lat}}$.}
    \label{fig:quasi_newton_calibration}
\end{figure}

The empirical trajectories highlight an intrinsic trade-off between geometric metric refinement and global parameter stability (Fig.~\ref{fig:quasi_newton_calibration}). The parameter estimation error $|\hat{\lambda} - \lambda_{\text{true}}|$ follows a characteristic U-shaped profile, decreasing steadily from $0.060$ at $T_{\text{L-BFGS}} = 10$ to its global minimum of $\approx 0.051$ at $T_{\text{L-BFGS}} = 40$ (Fig.~\ref{fig:quasi_newton_calibration}a). Beyond 80 iterations, unconstrained optimization begins overfitting the discrete stochastic fluctuations of the observed adjacency matrix, causing the parameter error to climb, while noticeably widening the 95\% confidence interval. Concurrently, the latent structure error $1 - \rho_{\text{lat}}$ exhibits a slow, monotonic decline across the entire budget (from $0.28$ down to $0.24$; Fig.~\ref{fig:quasi_newton_calibration}b). However, these marginal coordinate refinements beyond $T_{\text{L-BFGS}} \approx 40$ are rapidly outweighed by the escalating bias on the macroscopic mixing parameter. We therefore adopt $T_{\text{L-BFGS}} = 40$ as the default refinement budget, which acts as an optimal early-stopping point that minimizes parameter bias before noise overfitting sets in.

\section{Framework Robustness under Topological Noise and Edge Pruning}
\label{sec:perturbation_robustness}

To assess the structural stability and identifiability limits of the inference pipeline when observed networks deviate from idealized generative conditions, we conducted two complementary perturbation stress tests: isotropic background noise contamination and progressive random edge pruning.

To track performance across any continuous perturbation parameter $\xi$ (where $\xi = \epsilon$ denotes the injected background noise fraction and $\xi = p$ represents the random edge pruning ratio), we evaluate the degradation of the inferred model components relative to the unperturbed baseline at $\xi = 0$ via three normalized resilience indices:
\begin{equation}
\begin{split}
    \mathcal{R}_{\rho}(\xi) &= \frac{\langle \rho_{\text{lat}}(\xi) \rangle}{\langle \rho_{\text{lat}}(0) \rangle}, \\
    \mathcal{S}_{\lambda}(\xi) &= \max\left( 0, \, 1 - \frac{e_\lambda(\xi) - e_\lambda(0)}{1 - e_\lambda(0)} \right), \\
    \mathcal{S}_{r}(\xi) &= \max\left( 0, \, 1 - \frac{e_r(\xi) - e_r(0)}{1 - e_r(0)} \right),
\end{split}
\label{eq:unified_resilience_metrics}
\end{equation}
where $\rho_{\text{lat}}$ is the latent distance correlation, $e_\lambda(\xi) = \langle |\hat{\lambda}(\xi) - \lambda_{\text{true}}| \rangle$ is the absolute mixing error, and $e_r(\xi) = \langle |\hat{r}(\xi) - r_{\text{true}}| / r_{\text{true}} \rangle$ is the relative spatial reach error, with angle brackets denoting expectations over independent benchmark realizations. Under this formulation, each metric is bounded within $[0, 1]$, where unity represents complete structural stability and zero denotes the breakdown of parameter recovery.

In the first stress test, topological contamination is modeled by mixing the doubly-balanced connection propensity matrix $\widetilde{W}$ with an Erd\H{o}s--R\'enyi maximum-entropy floor:
\begin{equation}
    \widetilde{W}^{(\epsilon)}_{ij} = (1 - \epsilon) \widetilde{W}_{ij} + \epsilon W^{\text{noise}}_{ij}, \quad \text{with} \quad W^{\text{noise}}_{ij} = \frac{1 - \delta_{ij}}{N - 1},
\end{equation}
where $\epsilon \in [0.0, 0.5]$ parameterizes the injected noise intensity. Because the uniform noise term satisfies $\sum_{j} W^{\text{noise}}_{ij} = 1$, this convex combination strictly preserves total propensity mass ($\sum_{i,j} \widetilde{W}^{(\epsilon)}_{ij} = N$). Discrete observed networks are then conditionally sampled as $W^{\text{obs}}_{ij}(\epsilon) \sim \text{Poisson}(N \cdot \widetilde{W}^{(\epsilon)}_{ij})$, and model optimization is executed directly without providing prior knowledge of the contamination level $\epsilon$.

\begin{figure}[!htbp]
    \centering
    \includegraphics[width=\linewidth]{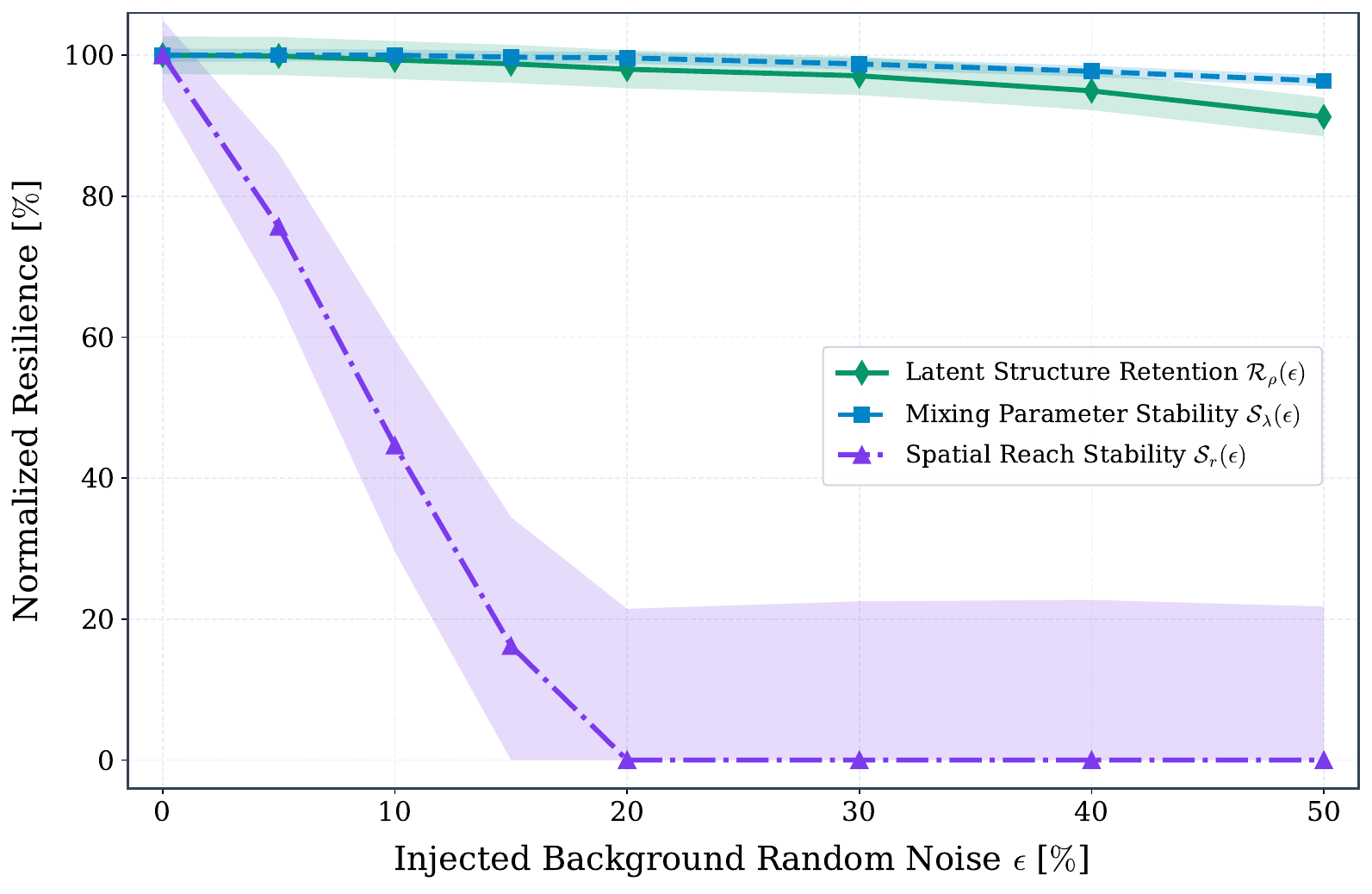}
    \caption{\textbf{Normalized framework resilience under uniform topological noise contamination.} 
    Degradation profiles evaluated as a function of injected background random noise $\epsilon \in [0.0, 0.50]$ averaged over 100 synthetic network realizations ($N \in [300, 2{,}000]$, $\lambda_{\text{true}} \sim \mathcal{U}(0, 1)$, $\ln r_{\text{true}}, \ln \sigma_{\text{true}} \sim \mathcal{U}(\ln 0.03, \ln 0.50)$). Plotted quantities show the normalized latent structure retention $\mathcal{R}_\rho(\epsilon)$ (green solid curve with diamonds), the macroscopic mixing parameter stability $\mathcal{S}_\lambda(\epsilon)$ (blue dashed curve with squares), and the physical spatial scale stability $\mathcal{S}_r(\epsilon)$ (purple dash-dotted curve with triangles). Shaded bands denote empirical 95\% confidence intervals.}
    \label{fig:epsilon_noise_robustness}
\end{figure}

The empirical trajectories under background noise reveal a distinct resilience hierarchy (Fig.~\ref{fig:epsilon_noise_robustness}). Macroscopic mixing stability $\mathcal{S}_\lambda$ remains above $99\%$ for $\epsilon \le 20\%$ and retains $\approx 96.5\%$ stability at $\epsilon = 50\%$, as isotropic noise affects spatial and functional channels uniformly without biasing macro-scale attribution. Similarly, latent structure retention $\mathcal{R}_\rho$ degrades gracefully ($\approx 98\%$ at $\epsilon = 20\%$ and $\approx 91\%$ at $\epsilon = 50\%$), demonstrating that spectral initialization acts as an effective low-pass filter against high-frequency topological perturbations. In contrast, physical reach stability $\mathcal{S}_r$ decays sharply ($\approx 45\%$ at $\epsilon = 10\%$) and collapses for $\epsilon \ge 20\%$, as non-local shortcuts flatten distance-decay gradients.

In the second perturbation stress test, simulating severe observational incompleteness via random edge pruning, we identify the set of active non-zero links $\mathcal{E} = \{(i, j) : i < j, \, W^{\text{obs}}_{ij} > 0\}$ and remove a uniform fraction $p \in [0.0, 0.95]$ of edges without replacement:
\begin{equation}
    W^{\text{pruned}}_{ij}(p) = 
    \begin{cases}
        0 & \text{if } (i, j) \in \mathcal{E}_{\text{pruned}}(p), \\
        W^{\text{obs}}_{ij} & \text{otherwise},
    \end{cases}
\end{equation}
where $|\mathcal{E}_{\text{pruned}}(p)| = \lfloor p \cdot |\mathcal{E}| \rfloor$. The full two-stage optimization pipeline is then re-executed directly on the sparsified matrix $W^{\text{pruned}}(p)$.

\begin{figure}[!htbp]
    \centering
    \includegraphics[width=\linewidth]{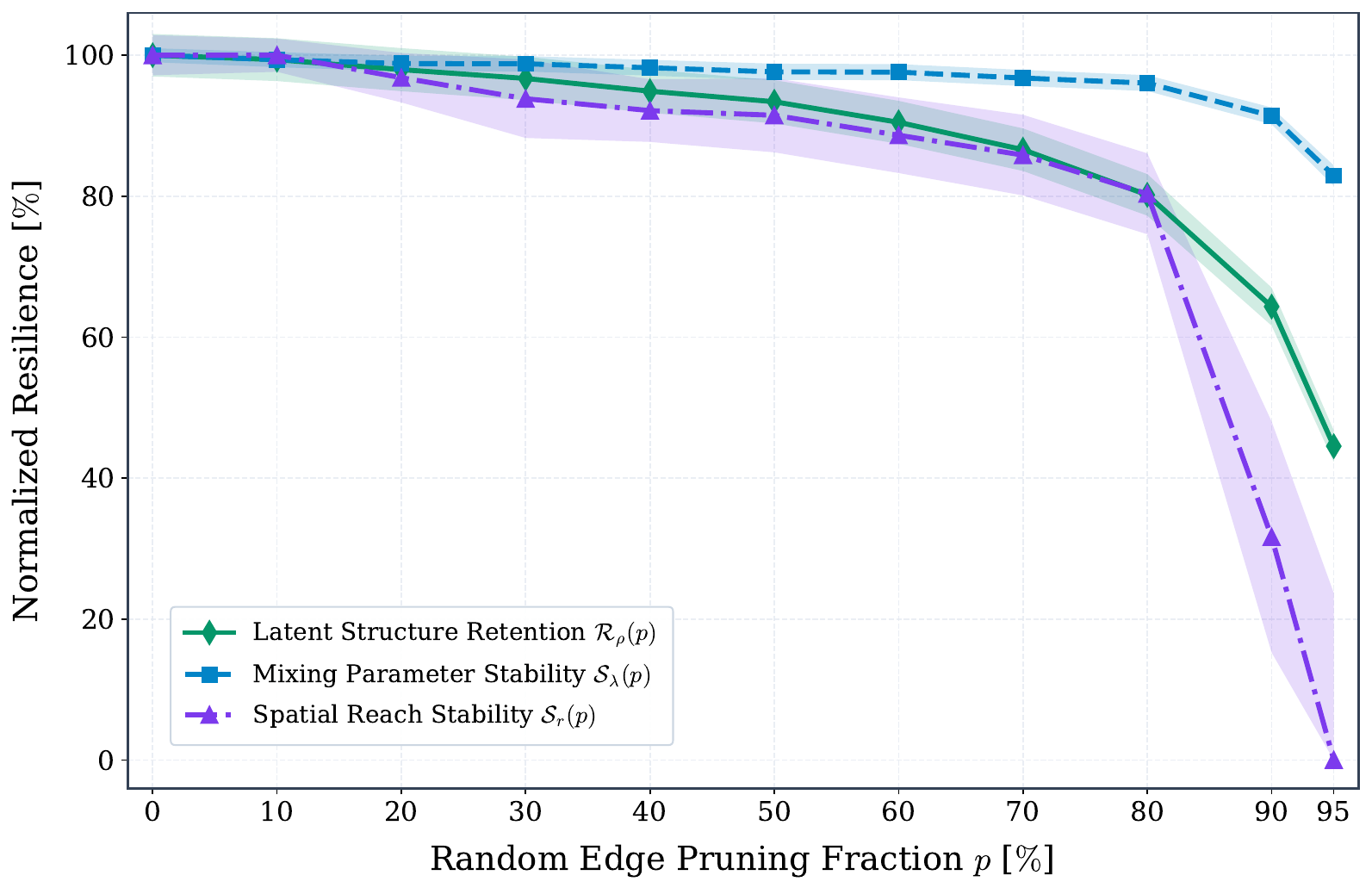}
    \caption{\textbf{Normalized framework resilience under random edge pruning.} 
    Degradation trajectories evaluated as a function of the random pruning fraction $p \in [0.0, 0.95]$ averaged over 100 synthetic network realizations ($N \in [300, 2{,}000]$, $\lambda_{\text{true}} \sim \mathcal{U}(0, 1)$, $\ln r_{\text{true}}, \ln \sigma_{\text{true}} \sim \mathcal{U}(\ln 0.03, \ln 0.50)$). Plotted metrics display the normalized latent structure retention $\mathcal{R}_\rho(p)$ (green solid curve with diamonds), the macroscopic mixing parameter stability $\mathcal{S}_\lambda(p)$ (blue dashed curve with squares), and the physical spatial scale stability $\mathcal{S}_r(p)$ (purple dash-dotted curve with triangles). Shaded bands denote empirical 95\% confidence intervals across ensemble realizations.}
    \label{fig:random_pruning_robustness}
\end{figure}

The degradation trajectories under edge pruning demonstrate that the global mixing share $\mathcal{S}_\lambda(p)$ remains nearly invariant up to $p = 80\%$ (Fig.~\ref{fig:random_pruning_robustness}). Simultaneously, latent structure retention $\mathcal{R}_\rho(p)$ and physical scale stability $\mathcal{S}_r(p)$ degrade in close alignment, retaining $\approx 80\%$ of their baseline accuracy up to $p = 80\%$. Beyond an $80\%$ pruning threshold, structural breakdown occurs: the physical reach scale collapses ($\mathcal{S}_r \approx 31\%$ at $p = 90\%$ and $\mathcal{S}_r \to 0$ at $p = 95\%$), while latent geometric retention falls to $\approx 44\%$.

Across both perturbation regimes, these stress tests confirm that while resolving continuous metric interaction ranges requires moderate network density and low background shortcuts, the macro-level socio-spatial allocation ($\lambda$) and the underlying latent geometry ($S$) remain robustly identifiable under severe data corruption and structural depletion.

\section{Robustness to Heavy-Tailed Distance Decay and Power-Law Kernels}
\label{sec:powerlaw_decay_analysis}

To assess whether our statistical inference pipeline generalizes beyond pure exponential distance decay, we evaluate its performance when spatial and relational affinities follow heavy-tailed power-law decay kernels. We couple physical and latent metric geometries via regularized, dimensionless Pareto-type decay kernels:
\begin{equation}
\begin{split}
    K^{\text{sp}}_{ij} &= \left(1 + \frac{D^{\text{phys}}_{ij}}{r_{\text{true}}}\right)^{-\alpha}, \\
    K^{\text{lat}}_{ij} &= \left(1 + \frac{D^{\text{lat}}_{ij}}{\sigma_{\text{true}}}\right)^{-\beta}, \quad \forall i \neq j,
\end{split}
\label{eq:powerlaw_kernels}
\end{equation}
where physical positions $\mathbf{x}_i \in [0, 1]^2$ determine the Euclidean distance $D^{\text{phys}}_{ij} = \|\mathbf{x}_i - \mathbf{x}_j\|_2$, and unobserved latent coordinates $\mathbf{s}_i \in [0, 1]^{d_s}$ are sampled from a Gaussian Mixture Model ($K \sim \mathcal{U}\{2, 8\}$ clusters with mixture weights $\boldsymbol{\pi} \sim \text{Dirichlet}(\mathbf{1}_K)$) yielding latent distances $D^{\text{lat}}_{ij} = \|\mathbf{s}_i - \mathbf{s}_j\|_2$. The additive unity regularizer prevents unphysical divergences at vanishing distances while preserving asymptotic power-law decay $K_{ij} \propto D^{-\gamma}$ at long ranges. Raw affinities $W^{\text{raw}}_{ij} = (1 - \lambda_{\text{true}}) K^{\text{sp}}_{ij} + \lambda_{\text{true}} K^{\text{lat}}_{ij}$ are row-balanced and Poisson-sampled as described in Appendix~\ref{app:benchmark_generation}.

\begin{figure}[!htbp]
    \centering
    \includegraphics[width=0.85\linewidth]{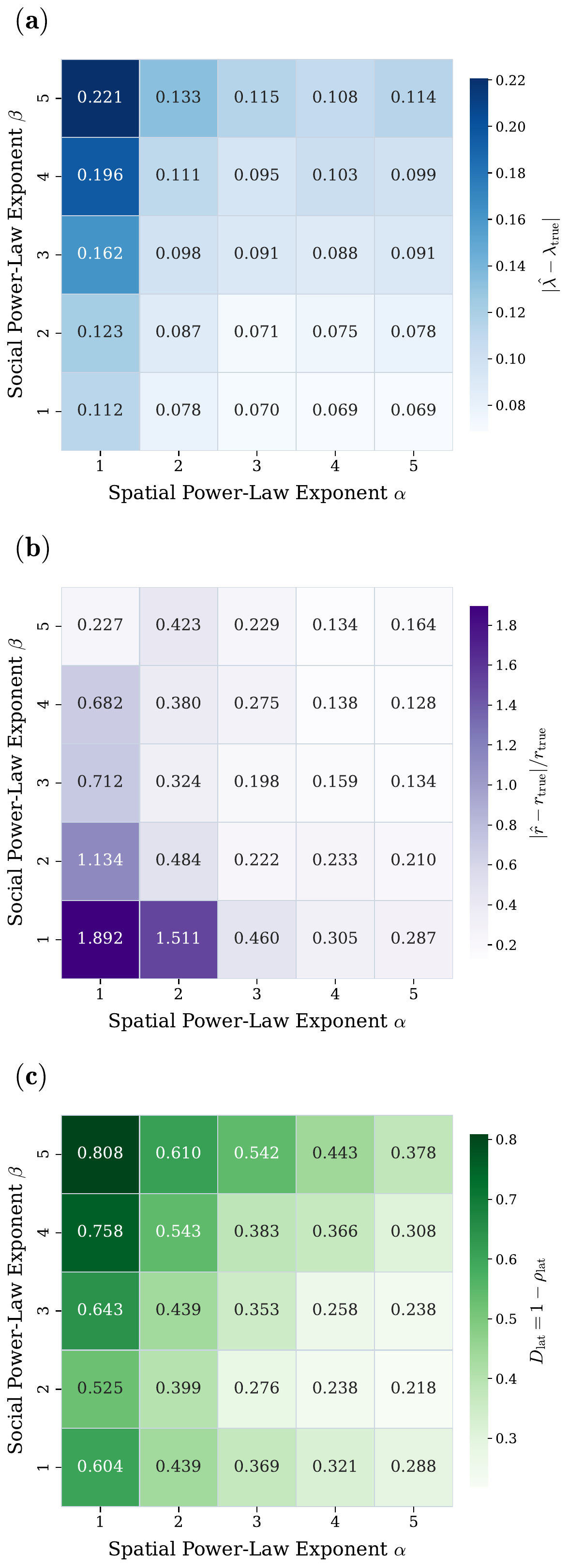}
    \caption{\textbf{Inference fidelity across dual power-law spatial and social decay regimes.} Heatmaps over a $5 \times 5$ grid of decay exponents $\alpha, \beta \in \{1, 2, 3, 4, 5\}$ ($M = 300$ realizations/cell, $N \in [100, 300]$, $\lambda_{\text{true}} \sim \mathcal{U}(0, 1)$, $\ln r_{\text{true}}, \ln \sigma_{\text{true}} \sim \mathcal{U}(\ln 0.03, \ln 0.50)$). \textbf{(a)} Mean parameter estimation error $|\hat{\lambda} - \lambda_{\text{true}}|$. \textbf{(b)} Relative reach scale error $|\hat{r} - r_{\text{true}}| / r_{\text{true}}$. \textbf{(c)} Latent metric space deficit $\mathcal{D}_{\text{lat}} = 1 - \rho_{\text{lat}}$.}
    \label{fig:powerlaw_heatmaps}
\end{figure}

We performed a two-dimensional grid sweep over $\alpha, \beta \in \{1, 2, 3, 4, 5\}$ across $M = 300$ synthetic network realizations per configuration, tracking the mixing error $e_\lambda = |\hat{\lambda} - \lambda_{\text{true}}|$, the relative spatial reach error $e_r = |\hat{r} - r_{\text{true}}| / r_{\text{true}}$, and the latent metric deficit $\mathcal{D}_{\text{lat}} = 1 - \rho_{\text{lat}}$ (Fig.~\ref{fig:powerlaw_heatmaps}).

The empirical error landscapes reveal three distinct operational regimes across the power-law exponents. First, the macroscopic coupling $\lambda$ is recovered with high fidelity across most of the configuration space (Fig.~\ref{fig:powerlaw_heatmaps}a), maintaining low error ($e_\lambda \approx 0.08$) under localized spatial regimes ($\alpha \ge 3$). Elevated bias is strictly confined to fat-tailed spatial interactions ($\alpha = 1$) paired with steep functional decay ($\beta \ge 4$, peaking at $e_\lambda \approx 0.22$), where diffuse physical links partially mimic non-local latent affinities. Second, physical reach recovery exhibits a sharp transition governed by $\alpha$ (Fig.~\ref{fig:powerlaw_heatmaps}b): flat likelihood gradients in scale-free regimes ($\alpha = 1$) inflate reach error up to $e_r \approx 1.89$, whereas spatial localization ($\alpha \ge 3$) restores steep metric gradients, bringing the error down to $\approx 0.20$. Third, while fat-tailed spatial links contaminate the residual operator at $\alpha = 1$ (elevating the metric deficit up to $\approx 0.81$; Fig.~\ref{fig:powerlaw_heatmaps}c), spatial confinement ($\alpha \ge 3$) enables spectral filtering to isolate community structure cleanly, reducing the deficit toward $\approx 0.22$ ($\rho_{\text{lat}} > 0.75$).

These results confirm that the inverse inference pipeline maintains robust identifiability across diverse algebraic distance decays, requiring only moderate spatial localization ($\alpha \ge 2$) to achieve high-precision geometric and parameter reconstruction.

\section{Latent Space Dimensionality and Model Selection}
\label{sec:latent_dimension_analysis}

While synthetic benchmarks evaluate a two-dimensional latent geometry ($d_s = 2$), empirical systems may exhibit unknown functional dimensions. To assess the robustness of the inference pipeline across varying latent complexity, we sampled ground-truth coordinates $S_{\text{true}} \in \mathbb{R}^{N \times d_s}$ from $d_s$-dimensional Gaussian Mixture Models with $K \sim \mathcal{U}\{2, 8\}$ clusters, scaled to $[0, 1]^{d_s}$. At the inference stage, the spectral initialization extracts a rank-$d_s$ truncated SVD on the non-spatial residual matrix, $R_{\text{init}} \approx U_{d_s} \Sigma_{d_s} U_{d_s}^T$, providing initial standardized embeddings $S^{(0)} = U_{d_s} \Sigma_{d_s}^{1/2} \in \mathbb{R}^{N \times d_s}$ before hybrid optimization. We systematically evaluated the framework across discrete dimensions $d_s \in [1, 8]$ over $M = 300$ synthetic network realizations per dimension.

\begin{figure}[!htbp]
    \centering
    \includegraphics[width=1\linewidth]{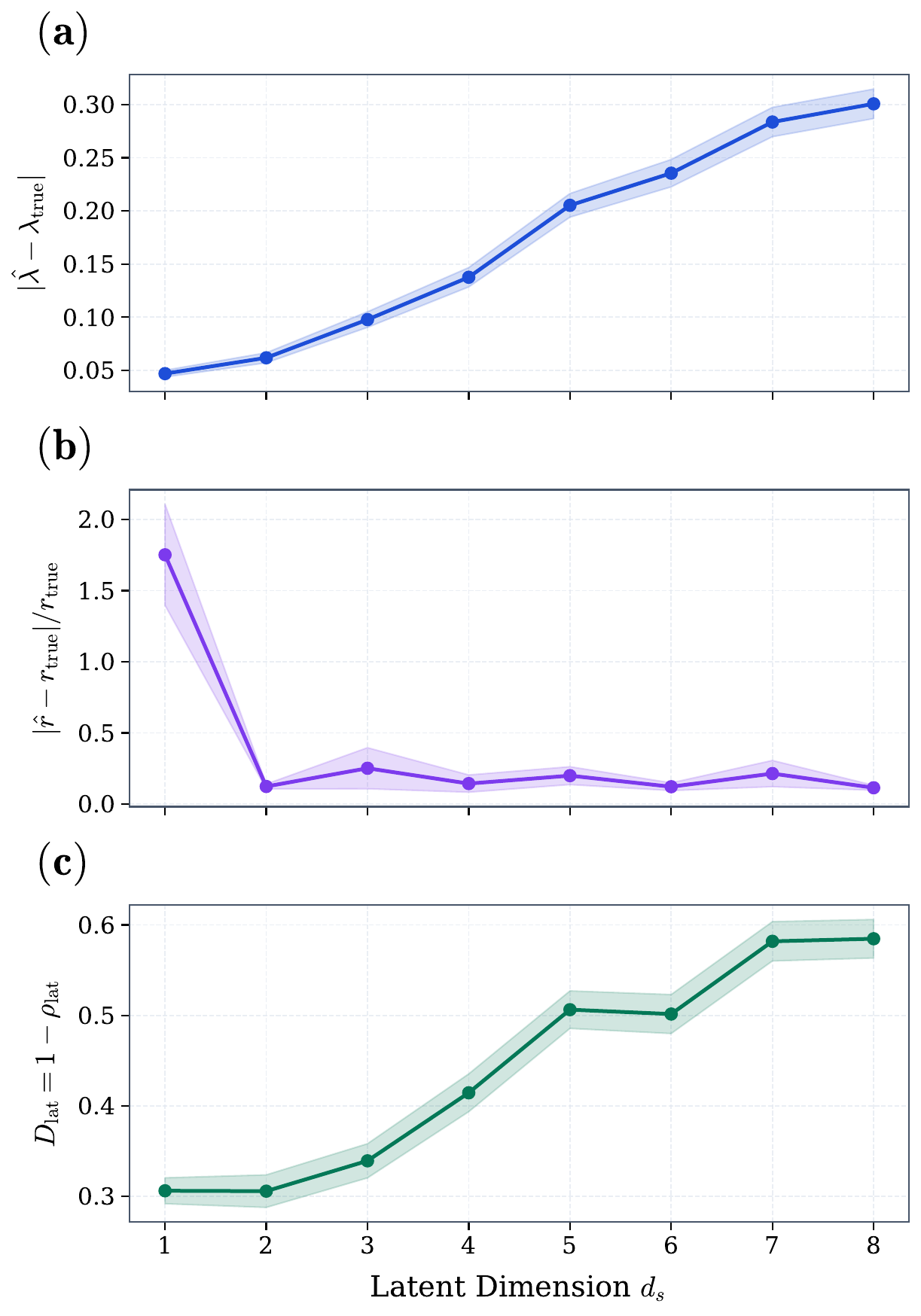}
    \caption{\textbf{Inference fidelity across latent social space dimensions $d_s \in [1, 8]$.} Performance evaluated across $M = 300$ realizations per dimension ($N \in [100, 300]$, $\lambda_{\text{true}} \in \mathcal{U}(0, 1)$, $\ln r_{\text{true}}, \ln \sigma_{\text{true}} \sim \mathcal{U}(\ln 0.03, \ln 0.50)$). Shaded bands denote $\pm 1$ standard error. \textbf{(a)} Mean parameter estimation error $|\hat{\lambda} - \lambda_{\text{true}}|$. \textbf{(b)} Mean relative physical reach error $|\hat{r} - r_{\text{true}}| / r_{\text{true}}$. \textbf{(c)} Latent metric space deficit $\mathcal{D}_{\text{lat}} = 1 - \rho_{\text{lat}}$.}
    \label{fig:dimension_robustness}
\end{figure}

The dimension sweep demonstrates how latent geometric complexity affects parameter identifiability and metric reconstruction (Fig.~\ref{fig:dimension_robustness}). 

For one-dimensional latent spaces ($d_s = 1$), the model accurately captures the global mixing share ($e_\lambda \approx 0.05$, Fig.~\ref{fig:dimension_robustness}a), but exhibits an elevated physical scale error ($e_r \approx 1.75$, Fig.~\ref{fig:dimension_robustness}b). This localized distortion stems from the rigid topological constraints of a 1D line, where a rank-1 spectral residual embedding partially conflates linear latent sorting with continuous geographic distance decay. Once the latent space spans $d_s \ge 2$, however, this artifact vanishes: the physical reach scale decouples cleanly, maintaining a low and stable error floor ($e_r \approx 0.12\text{--}0.20$) across all higher dimensions.

Concurrently, as the latent dimensionality grows from $d_s = 2$ to $d_s = 8$, both the mixing parameter error and the metric deficit increase monotonically ($e_\lambda$ climbs from $0.06$ to $0.30$, and $\mathcal{D}_{\text{lat}}$ rises from $0.30$ to $0.58$, Fig.~\ref{fig:dimension_robustness}a,c). This trend illustrates the standard statistical costs of high-dimensional embedding: Euclidean distance concentration flattens affinity contrasts, while optimizing an expanding budget of $N \times d_s$ continuous parameters against discrete Poisson observations lowers statistical power. Nevertheless, recovery remains sharp ($e_\lambda \le 0.10$, $\mathcal{D}_{\text{lat}} \le 0.35$) across the low-dimensional domain ($d_s \le 3$) that typically characterizes real-world relational systems.

In empirical applications where the latent space dimension is unknown, the optimal embedding dimension can be determined either a priori via spectral scree analysis of the non-spatial residual matrix $R_{\text{init}}$ (locating the spectral gap before singular values enter the Marchenko--Pastur noise floor) or through edge-masking cross-validation by selecting the dimension $d_s^*$ that maximizes out-of-sample Poisson predictive likelihood.

\section{Likelihood Optimization vs. Deep Graph Neural Networks}
\label{sec:gnn_vs_mle_benchmark}

To evaluate the inference fidelity and parameter identifiability of our framework against standard geometric deep learning approaches, we benchmarked the hybrid maximum likelihood pipeline against a specialized Dual-Kernel Spatial Graph Neural Network (GNN). Here, we compare the training dynamics, sample efficiency, and asymptotic parameter estimation precision of both paradigms on synthetic spatial network ensembles.

\begin{figure}[!htbp]
    \centering
    \includegraphics[width=0.85\columnwidth]{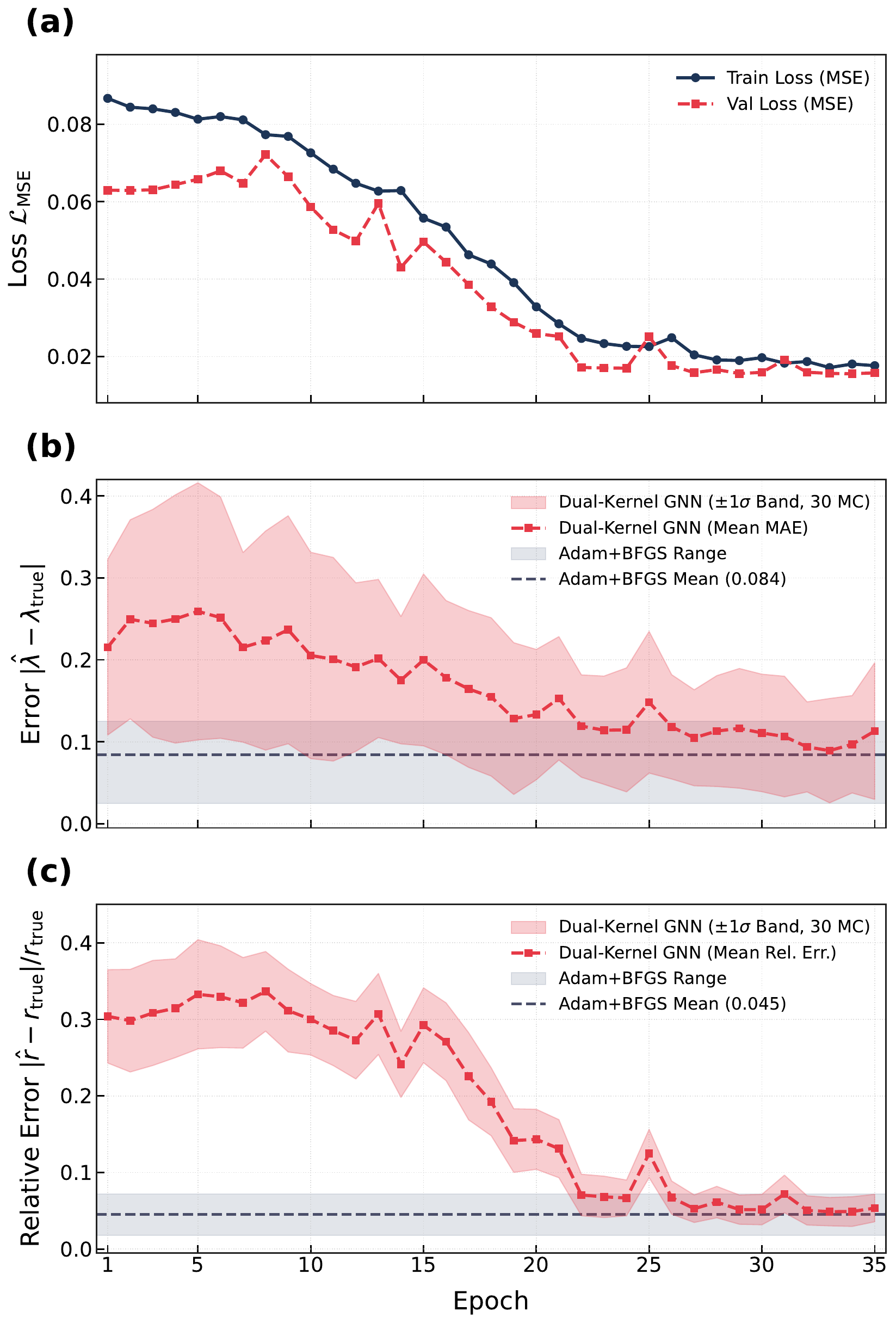}
    \caption{\textbf{Learning dynamics and convergence benchmark of the Dual-Kernel Spatial GNN against analytical likelihood estimation.} 
    \textbf{(a)} Training (blue solid line) and validation (red dashed line) mean squared error loss trajectories $\mathcal{L}_{\text{MSE}}$ over 35 epochs. 
    \textbf{(b)} Mean absolute error on the spatio-functional coupling parameter $e_\lambda = |\hat{\lambda} - \lambda_{\text{true}}|$ across training epochs. Red squares denote GNN mean performance, with the shaded pink envelope showing the $\pm 1\sigma$ band across 30 independent Monte Carlo runs. The horizontal dashed line and gray shaded region denote the mean ($0.084$) and empirical variation range of the analytical Adam+BFGS framework. 
    \textbf{(c)} Relative estimation error on the physical interaction scale $e_r = |\hat{r} - r_{\text{true}}| / r_{\text{true}}$. Blue-green squares and shaded envelope indicate GNN trajectory ($\pm 1\sigma$, 30 Monte Carlo runs), compared against the analytical Adam+BFGS benchmark (dashed line at mean $0.045$, with gray empirical range).}
    \label{fig:gnn_convergence_benchmark}
\end{figure}

The baseline GNN was engineered to process spatially embedded, weighted graphs by explicitly integrating metric geometry with observed topology. For each node $i$, the model constructs an input feature vector concatenating physical coordinates $\mathbf{x}_i \in [0, 1]^2$, normalized degree $d_i / N$, and Euclidean distance to the spatial centroid $\|\mathbf{x}_i - \bar{\mathbf{x}}\|_2$. Corresponding edge attributes combine the Euclidean physical distance $D_{ij}^{\text{phys}}$, observed connection weight $W_{ij}^{\text{obs}}$, binary link existence $\mathbb{I}(W_{ij}^{\text{obs}} > 0)$, and prior spatial kernel decay $\exp(-D_{ij}^{\text{phys}} / 0.25)$. Across $L = 3$ spatial message passing layers, directional messages are aggregated over an augmented adjacency combining observed links with local metric proximity ($D_{ij}^{\text{phys}} < 0.15$), followed by residual updates and Layer Normalization. Graph-level representations are extracted via multi-channel pooling combining global mean, global maximum, and degree-weighted node embeddings, feeding into a multi-layer perceptron head designed to jointly predict the spat-functional coupling $\hat{\lambda}$ and physical interaction reach $\hat{r}$. The network is trained end-to-end by minimizing the mean squared error on the spatio-functional coupling parameter:
\begin{equation}
    \mathcal{L}_{\text{MSE}} = \frac{1}{|B|} \sum_{b \in B} \left(\hat{\lambda}_b - \lambda_{b, \text{true}}\right)^2,
    \label{eq:gnn_mse_loss}
\end{equation}
where $B$ denotes the training batch. The neural network was trained over 35 epochs via AdamW with cosine annealing ($\eta \in [5 \times 10^{-5}, 10^{-3}]$) on $N_{\text{train}} = 360$ synthetic networks ($N \in [100, 300]$, $\lambda_{\text{true}} \sim \mathcal{U}(0, 1)$, $\ln r_{\text{true}}, \ln \sigma_{\text{true}} \sim \mathcal{U}(\ln 0.03, \ln 0.50)$), monitored at each epoch on $N_{\text{val}} = 60$ validation graphs, and evaluated on an independent clean test ensemble of $N_{\text{test}} = 40$ networks across 30 independent Monte Carlo trials.

Conversely, our analytical likelihood pipeline requires zero offline training corpora or parameter supervision. Instead of regressing against ground-truth labels, each network instance is solved independently by minimizing the regularized Poisson negative log-likelihood directly on observed connectivity:
\begin{equation}
\begin{split}
    \mathcal{L}(\theta_\lambda, \ln r, \mathbf{S}) &= \sum_{i < j} \left( W_{ij}^{\text{pred}} - W_{ij}^{\text{obs}} \ln W_{ij}^{\text{pred}} \right) \\
    &\quad + \frac{1}{2}\left(\frac{\ln r - \ln r_0}{2}\right)^2,
\end{split}
\end{equation}
initialized via rank-2 truncated SVD spectral decomposition on the non-spatial residual matrix $R^+$ and optimized through 40 Adam warmup iterations followed by 40 quasi-Newton L-BFGS refinement steps.

Tracking convergence dynamics across the 35 training epochs reveals distinct learning trajectories and highlights the inductive advantages of likelihood-based inference (Fig.~\ref{fig:gnn_convergence_benchmark}). As shown in Fig.~\ref{fig:gnn_convergence_benchmark}a, both training and validation losses decrease monotonically from $\mathcal{L}_{\text{MSE}} \approx 0.08$ to a stable plateau around $\approx 0.02$, confirming that the spatial message passing layers successfully extract coordinated representations from embedded graph topologies without overfitting.

This loss contraction translates into a progressive refinement of the spatio-functional coupling parameter predictions (Fig.~\ref{fig:gnn_convergence_benchmark}b). The GNN mean absolute error on the macroscopic coupling $\lambda$ decreases monotonically from an initial $e_\lambda \approx 0.25$ down to approximately $0.10$ by epoch 35, approaching the performance floor achieved zero-shot by our unsupervised Adam+BFGS pipeline (mean $e_\lambda = 0.084$, gray shaded interval).

The contrast is even more pronounced when resolving the continuous physical interaction reach $r$ (Fig.~\ref{fig:gnn_convergence_benchmark}c). While the GNN relative error improves steadily from $e_r \approx 0.35$ to an asymptotic floor of $e_r \approx 0.09$, the hybrid analytical optimizer maintains a markedly superior resolution, converging to a mean relative error of $0.045$ with minimal variance.

\section{Network Size Scaling and Computational Runtime}
\label{sec:system_size_scaling}

\begin{figure}[!htbp]
    \centering
    \includegraphics[width=\linewidth]{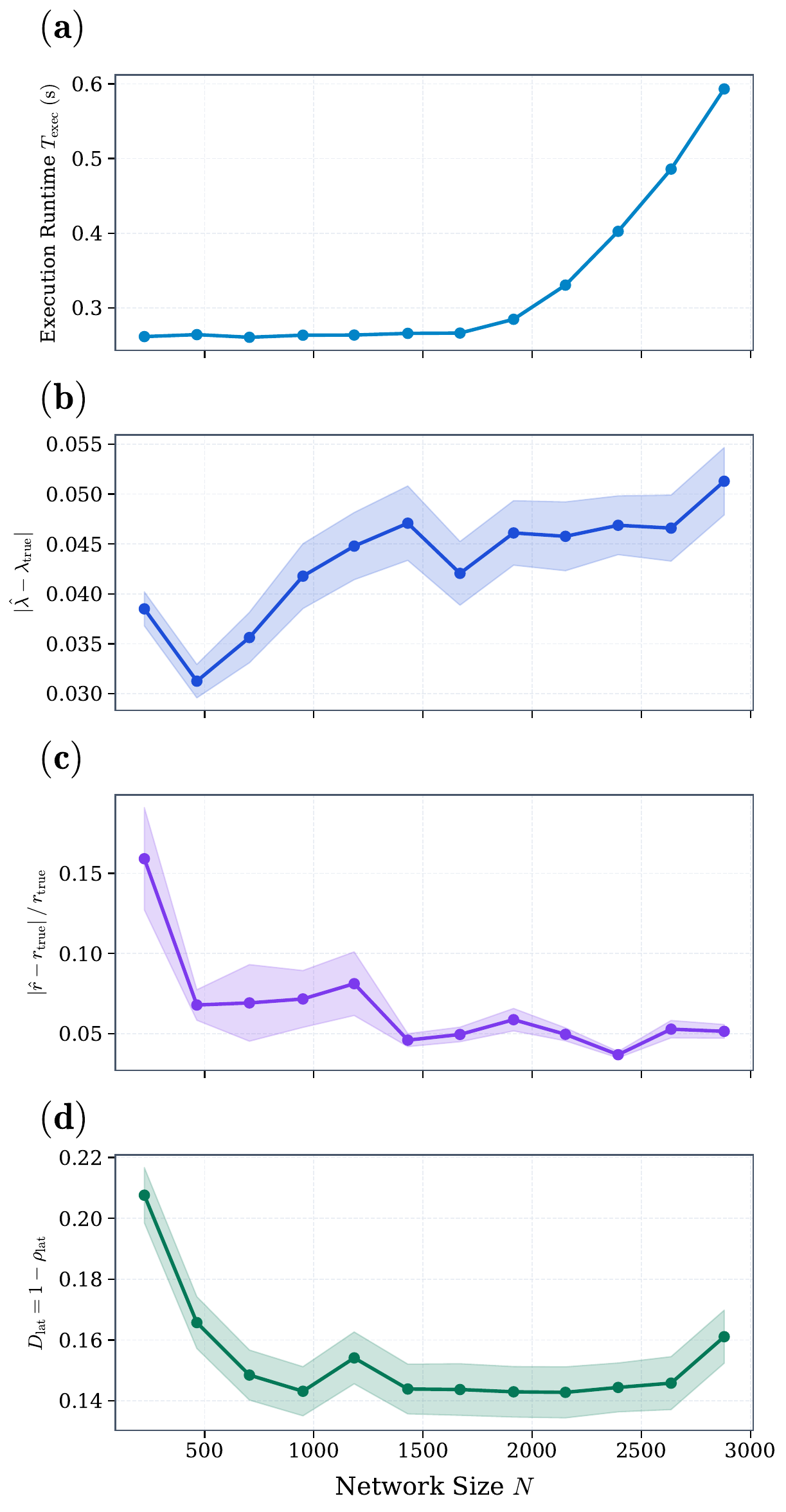}
    \caption{\textbf{Computational scaling and inference fidelity across network sizes $N \in [100, 3{,}000]$.} Statistics evaluated over $M = 10{,}000$ synthetic network realizations partitioned into 12 uniform bins across $N$ ($\approx 800\text{--}860$ realizations per bin). Points represent bin averages; shaded regions indicate $\pm 1$ standard error of the mean. \textbf{(a)} Mean GPU execution runtime $T_{\text{exec}}$. \textbf{(b)} Mean parameter estimation error $|\hat{\lambda} - \lambda_{\text{true}}|$. \textbf{(c)} Mean relative physical reach error $|\hat{r} - r_{\text{true}}| / r_{\text{true}}$. \textbf{(d)} Latent metric space deficit $\mathcal{D}_{\text{lat}} = 1 - \rho_{\text{lat}}$.}
    \label{fig:scaling_analysis}
\end{figure}

To evaluate asymptotic statistical consistency and benchmark computational throughput as a function of system size, we executed a large-scale scaling analysis across synthetic network ensembles ranging from $N = 100$ to $N = 3{,}000$ nodes. We evaluated $M = 10{,}000$ independent synthetic network realizations sampled uniformly across graph sizes, drawing ground-truth parameters from uninformative priors ($\lambda_{\text{true}} \sim \mathcal{U}(0, 1)$, $\ln r_{\text{true}}, \ln \sigma_{\text{true}} \sim \mathcal{U}(\ln 0.03, \ln 0.50)$). The ensemble was partitioned into 12 contiguous bins across $N$ to track the scaling of execution runtime $T_{\text{exec}}$, absolute mixing error $e_\lambda = |\hat{\lambda} - \lambda_{\text{true}}|$, relative reach scale error $e_r = |\hat{r} - r_{\text{true}}| / r_{\text{true}}$, and latent metric deficit $\mathcal{D}_{\text{lat}} = 1 - \rho_{\text{lat}}$ (Fig.~\ref{fig:scaling_analysis}).

The empirical trajectories confirm both sub-second computational efficiency and asymptotic statistical consistency. On GPU architectures, inference generally remains below $0.60\text{ s}$ across all configurations (Fig.~\ref{fig:scaling_analysis}a), while the macroscopic mixing error $|\hat{\lambda} - \lambda_{\text{true}}|$ remains stably bounded around $\approx 0.04$ (Fig.~\ref{fig:scaling_analysis}b). While small systems ($N \le 300$) exhibit finite-size resolution limits ($e_r \approx 0.16$, $\mathcal{D}_{\text{lat}} \approx 0.21$), larger networks ($N \ge 500$) provide sufficient pairwise statistics to collapse the spatial reach error down to $e_r \approx 0.05$ and stabilize the latent metric deficit at $\mathcal{D}_{\text{lat}} \approx 0.14$ ($\rho_{\text{lat}} \ge 0.85$, Fig.~\ref{fig:scaling_analysis}c,d).

Nevertheless, exact full-batch evaluation incurs an $\mathcal{O}(N^2)$ computational and memory footprint that limits direct execution on massive networks. To scale beyond this regime to graphs with $N \ge 10^4$ nodes, two complementary algorithmic strategies can be deployed. 

First, streaming chunk-wise normalization eliminates the dense $N \times N$ allocation required to compute the expected nodal strengths $s_i = \sum_{j=1}^N W_{ij}^{\text{raw}}$. By retaining node coordinates in $\mathcal{O}(N)$ memory ($X \in \mathbb{R}^{N \times 2}$ and $S \in \mathbb{R}^{N \times d_s}$) and evaluating interactions in streaming row-batches of size $b \ll N$ (e.g., $b = 2{,}500$), transient pairwise distance blocks $(D_{uv})^2 = \|x_u\|^2 + \|x_v\|^2 - 2 x_u^T x_v$ and affinities $W^{\text{raw}}_{B, :}$ are accumulated directly into a row-sum buffer and immediately deallocated. This bounds total GPU memory consumption below $150\text{ MB}$, irrespective of network size $N$.

Second, induced subgraph invariance enables scalable parameter estimation via composite likelihood ensembling. For an induced subgraph $\mathcal{V}_{\text{sub}} \subset \mathcal{V}$ of $n \ll N$ uniformly sampled nodes, metric distances $\|x_u - x_v\|_2$ and $\|s_u - s_v\|_2$ remain strictly invariant to total system volume $N$. Because the coupling $\lambda$ and physical scale $r$ act as intensive parameters governing continuous distance-decay rates, evaluating the restricted Poisson log-likelihood on the induced subgraph:
\begin{equation}
\begin{split}
    \mathcal{L}_{\text{sub}}(\lambda, r, S_{\text{sub}}) &= \sum_{u < v \in \mathcal{V}_{\text{sub}}} \Bigl[ \Lambda_{uv}(\lambda, r, S_{\text{sub}}) \\
    &\quad - W^{\text{obs}}_{uv} \ln \Lambda_{uv}(\lambda, r, S_{\text{sub}}) \Bigr],
\end{split}
\label{eq:subgraph_likelihood}
\end{equation}
defines a valid composite marginal likelihood, where local propensity normalization within $\mathcal{V}_{\text{sub}}$ consistently preserves relative affinity contrasts. This guarantees that optimizing $\mathcal{L}_{\text{sub}}$ across independent subgraphs yields statistically consistent estimators of $(\lambda, r)$ while bypassing the $\mathcal{O}(N^2)$ global memory allocation.

To ensure numerical stability and minimize sample variance, local inference on each induced subgraph executes the two-stage hybrid protocol (rank-$d_s$ truncated SVD on spatial residuals $R^+ = \max(0, W_{\text{sub}} - W_{\text{sp}0})$, Adam exploration, and L-BFGS quasi-Newton polish). Final parameter estimates are aggregated across an ensemble of $K$ independent subgraphs ($K = 40$):
\begin{equation}
    \bar{\lambda} = \frac{1}{K} \sum_{k=1}^K \hat{\lambda}_k, \qquad \bar{r} = \frac{1}{K} \sum_{k=1}^K \hat{r}_k,
\end{equation}
yielding a Monte Carlo standard error that suppresses estimator variance as $\sigma_{\bar{\lambda}} = \sigma_\lambda / \sqrt{K}$ while permitting trivial parallelization across GPU streams and CPU worker pools.

\begin{figure}[!htbp]
    \centering
    \includegraphics[width=\linewidth]{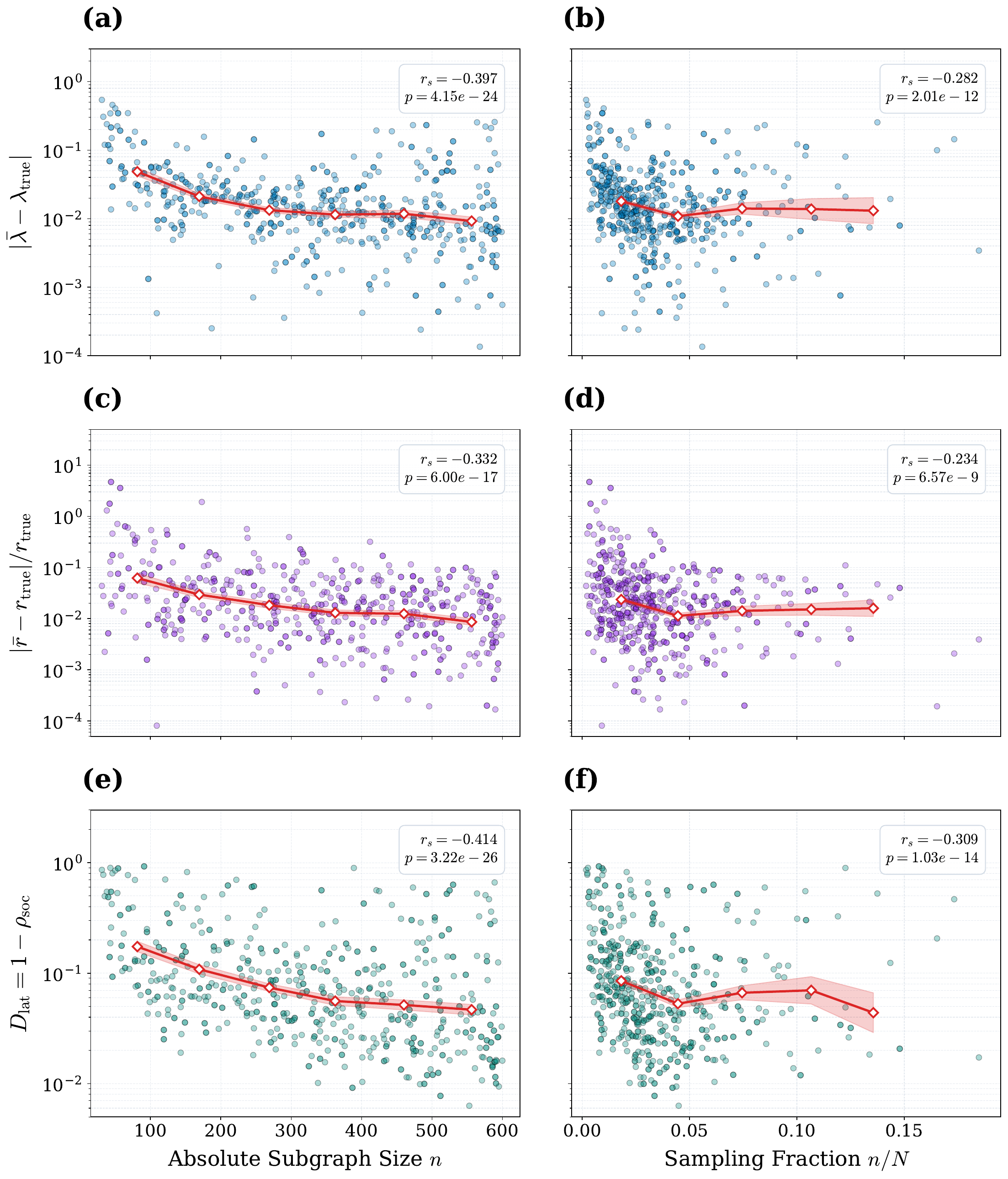}
    \caption{\textbf{Induced subgraph scaling and asymptotic parameter recovery across 600 synthetic benchmark realizations.} Left panels \textbf{(a, c, e)} display estimation errors against absolute subgraph size $n \in [50, 600]$, where red solid curves with open diamonds and shaded envelopes denote binned geometric means and associated standard errors of the mean, respectively. Right panels \textbf{(b, d, f)} plot the same realizations against the relative sampling fraction $n/N$. \textbf{(a, b)} Absolute mixing error $|\bar{\lambda} - \lambda_{\text{true}}|$. \textbf{(c, d)} Relative physical reach error $|\bar{r} - r_{\text{true}}| / r_{\text{true}}$. \textbf{(e, f)} Latent metric space deficit $\mathcal{D}_{\text{lat}} = 1 - \rho_{\text{lat}}$. Insets report Spearman rank correlation coefficients $r_s$ and associated two-sided $p$-values.}
    \label{fig:subgraph_scaling_errors}
\end{figure}

Empirical validation across $N_{\text{exp}} = 600$ benchmark realizations confirms the theoretical guarantees of the induced subgraph formulation (Fig.~\ref{fig:subgraph_scaling_errors}). In particular, comparing estimation errors against the relative sampling fraction $n/N$ versus the absolute subgraph size $n$ reveals that $n/N$ acts as a spurious variable: plotting error trajectories against $n/N$ produces dispersed, uninformative profiles (Fig.~\ref{fig:subgraph_scaling_errors}b,d,f). In contrast, aligning performance against the absolute sample size $n$ unveils strong, statistically robust power-law decays across all three metrics (Fig.~\ref{fig:subgraph_scaling_errors}a,c,e), confirmed by significant Spearman rank correlations ($r_s = -0.397$ for mixing error, $r_s = -0.332$ for physical reach scale, and $r_s = -0.414$ for latent metric deficit; $p < 10^{-16}$ across all observables).

For moderate local subgraphs of $n \approx 300$ nodes, the average mixing parameter error contracts to $|\bar{\lambda} - \lambda_{\text{true}}| \approx 0.01$, while larger subgraphs ($n \ge 500$) provide sufficient pairwise statistics to collapse the physical reach error below $|\bar{r} - r_{\text{true}}| / r_{\text{true}} \le 0.01$ and stabilize the latent distance deficit at $1 - \rho_{\text{lat}} \le 0.05$ (corresponding to distance correlations $\rho_{\text{lat}} \ge 0.95$). By replacing full-graph optimization with local ensemble sampling, runtime complexity drops from $\mathcal{O}(N^2 T)$ to $\mathcal{O}(K n^2 T)$ and memory footprints collapse from $\mathcal{O}(N^2)$ to $\mathcal{O}(N + n^2)$, enabling rigorous socio-spatial inverse inference on massive real-world networks with $N \ge 10^5$ nodes.

\newpage

\bibliography{references}

\end{document}